\documentclass[12pt]{article}      
\usepackage{latexsym,amssymb}      
\newcommand{\mathsc}{\scriptscriptstyle\rm}
\newcommand{\preprint}[1]{\thispagestyle{empty}~\newline
 \vspace*{-22.65mm}\begin{flushright}\begin{tabular}{l} #1
 \end{tabular}\end{flushright}\vspace{1cm}}
\catcode`\@=11 \@addtoreset{equation}{section}\catcode`\@=11

\renewcommand{\subsection}[1]{\indent\par{\it #1.--- }}
\def\sso{\$}
\def\bibtit#1{{\sl #1}}
\renewcommand{\Re}{\mathbb{R}}

\begin{document}

\def\fecha{2 March 1999}
\def\titulo{Quantum evolution in spacetime foam}
\def\autor{Luis J. Garay}
\def\direccion{Instituto de Matem\'{a}ticas y F\'{\i}sica Fundamental,
 CSIC, C/ Serrano  121, 28006 Madrid, Spain}

\markright{\titulo}

\preprint{gr-qc/9911002}
\vspace*{-10mm}
\begin{center}
 {\Large\bf \titulo}\\[5mm]
 {\autor}\\
 {\small\it \direccion}\\
 {\small \fecha}
\end{center}

\begin{abstract}
In this work, I review some aspects concerning the evolution of
quantum low-energy fields in a foamlike spacetime, with involved
topology at the Planck scale but with a smooth metric structure
at large length scales, as follows. Quantum gravitational
fluctuations may induce a minimum length thus introducing an
additional source of uncertainty in physics. The existence of
this resolution limit casts doubts on the metric structure of
spacetime at the Planck scale and opens a doorway to nontrivial
topologies, which may dominate Planck scale physics. This
foamlike structure of spacetime may show up in low-energy
physics through loss of quantum coherence and mode-dependent
energy shifts, for instance, which might be observable.
Spacetime foam introduces nonlocal interactions that can be
modeled by a quantum bath, and low-energy fields evolve
according to a master equation that displays such effects.
Similar laws are also obtained for quantum mechanical systems
evolving according to good real clocks, although the underlying
Hamiltonian structure in this case establishes serious
differences among both scenarios. {\it Contents.--- } Quantum
fluctuations of the gravitational field; Spacetime foam; Loss of
quantum coherence; Quantum bath; Low-energy effective evolution;
Real clocks; Conclusions.\\

\hspace*{1mm}\hfill \textit{Int. J. Mod. Phys. A14 (1999) 4079--4120}
\end{abstract}


\section{Quantum fluctuations of the gravitational field}
\indent

Gravity deals with the frame in which everything takes place,
i.e., with spacetime. We are used to putting everything into
spacetime, so that we can name and handle events. General
relativity made spacetime dynamical but the relations between
different events were still sharply defined. Because of quantum
mechanics, in such a dynamical frame, objects became fuzzy;
exact locations were substituted by probability amplitudes of
finding an object in a given region of space at a given instant
of time. Spacetime undergoes the quantum fluctuations of the
other interactions and, even more, introduces its own
fluctuations, thus becoming an active agent in the theory. The
quantum theory of gravity suffers from problems (see, e.g. Refs.
\cite{is93,is97}) that have remained unsolved for many years and
that are originated, in part, in this lack of a fixed immutable
spacetime background.

A quantum uncertainty in the position of a particle implies an
uncertainty in its momentum and, therefore, due to the
gravity-energy universal interaction, would also imply an
uncertainty in the geometry, which in turn would introduce an
additional uncertainty in position of the particle. The geometry
would thus be subject to quantum fluctuations that would
constitute the spacetime foam and that should be of the same
order as the geometry itself at the Planck scale. This would
give rise to a minimum length \cite{ga95} beyond which the
geometrical properties of spacetime would be lost, while on
larger scales it would look smooth and with a well-defined
metric structure. The key ingredients for the appearance of this
minimum length are quantum mechanics, special relativity, which
is essential for the unification of all kinds of energy via the
finiteness of the speed of light, and a theory of gravity, i.e.,
a theory that accounts for the active response of spacetime to
the presence of energy (general relativity, Sakharov's
elasticity \cite{sa67,mt73}, strings\ldots). Thus, the existence
of a lower bound to any output of a position measurement, seems
to be a model-independent feature of quantum gravity. In fact,
different approaches to this theory lead to this result
\cite{ga95}.

Planck length $\ell_*$ might play a role analogous to the speed
of light in special relativity. In this theory, there is no
physics beyond this speed limit and its existence may be
inferred through the relativistic corrections to the Newtonian
behavior. This would mean that a quantum theory of gravity could
be constructed only on ``this side of Planck's border'' as
pointed out by Markov \cite{ma80,ma81} (as quoted in Ref.
\cite{bt88}). In fact, the analogy between quantum gravity and
special relativity seems to be quite close: in the latter you
can accelerate forever even though you will never reach the
speed of light; in the former, given a coordinate frame, you can
reduce the coordinate distance between two events as much as you
want even though the proper distance between them will never
decrease beyond Planck length (see Ref. \cite{ga95}, and
references therein). This uncertainty relation $\Delta x\geq
\ell_*$ also bears a close resemblance to the role of $\hbar$ in
quantum mechanics: no matter which variables are used, it is not
possible to have an action $S$ smaller than $\hbar$
\cite{me91,mensky98,me92}.

Based on the work by Bohr and Rosenfeld
\cite{br33,ro63,br50,ro55} (see e.g. Ref. \cite{he54}) for the
electromagnetic field, Peres and Rosen \cite{pr60} and then
DeWitt \cite{de62} carefully analyzed the measurement of the
gravitational field and the possible sources of uncertainty (see
also Refs. \cite{bt88,re58,bt82,tr85}). Their analysis was
carried out in the weak-field approximation (the magnitude of
the Riemann tensor remains small for any finite domain) although
the features under study can be seen to have more fundamental
significance. This approximation imposes a limitation on the
bodies that generate and suffer the gravitational field, which
does not appear in the case of an electromagnetic field. The
main reason for this is that, in this case, the relevant
quantity that is involved in the uncertainty relations is the
ratio between the charge and the mass of the test body, and this
quantity can be made arbitrarily small. This is certainly not
the case for gravitational interactions, since the equivalence
principle precisely fixes the corresponding ratio between
gravitational mass and inertial mass, and therefore it is not
possible to make it arbitrarily small. Let us go into more
detail in the comparison between the electromagnetic and the
gravitational fields as far as uncertainties in the measurement
are concerned and see how it naturally leads to a minimum volume
of the measurement domain.

The measurement of the gravitational field can be studied from
the point of view of continuous measurements
\cite{me91,mensky98,me92}, which we briefly summarize in what
follows (throughout this work we set $\hbar=c=1$, so that the
only dimensional constant is Planck's length
$\ell_*=\sqrt{\mbox{\small G}}$).

\subsection{Continuous measurements}
Assume that we continuously measure an observable $Q$, with\-in
the framework of ordinary quantum mechanics. Let us call $\Delta
q$ the uncertainty of our measurement device. This means that,
as a result of our measurement, we will obtain an output
$\alpha$ that will consist of the result $q(t)$ and any other
within the range $(q-\Delta q, q+\Delta q)$. The probability
amplitude for an output $\alpha$ can be written in terms of path
integrals \cite{me91,mensky98} $A[\alpha]=\int_\alpha {\cal D} x
e^{iS}$, where $\alpha$ denotes not only the output but also the
set of trajectories in configuration space that lead to it. For
a given uncertainty $\Delta q$, the set $\alpha$ is fully
characterized by its central value $q$. We are particularly
interested in studying the shape of the probability amplitude
$A$. More precisely, we will pay special attention to its width
$\Delta
\alpha$ \cite{me91,mensky98,me92}.

There are two different regimes of measurement, classical and
quantum, depending on whether the uncertainty of the measuring
device is large or small. The classical regime of measurement
will be accomplished if $\Delta q$ is large enough. In this
regime, the width of the probability amplitude $\Delta \alpha$
can be seen to be proportional to the uncertainty $\Delta q$.
Also, the uncertainty on the action can be estimated to be
$\Delta S\gtrsim 1$. The quantum regime of measurement occurs
when $\Delta q$ is very small. Now the width of the probability
amplitude is $\Delta
\alpha\sim 1/\Delta q$. The uncertainty in the action is also
greater than unity in this case.

Thus, in any regime of measurement, the action uncertainty will
be greater than unity. In view of this discussion, the width
$\Delta \alpha$ of the probability amplitude will achieve its
minimum value, i.e., the measurement will be optimized, for
uncertainties in the measurement device $\Delta q$ that are
neither too large nor too small. When this minimum nonvanishing
value is achieved, the uncertainty in the action is also
minimized and set equal to one. The limitation on the accuracy
of any continuous measurement is, of course, an expression of
Heisenberg's uncertainty principle. Since we are talking about
measuring trajectories in some sense, a resolution limit should
appear, expressing the fact that position and momentum cannot be
measured simultaneously with infinite accuracy. In the classical
regime of measurement, the accuracy is limited by the intrinsic
uncertainty of the measuring device. On the other hand, when
very accurate devices are employed, quantum fluctuations of the
measuring apparatus affect the measured system and the final
accuracy is also affected. The maximum accuracy is obtained when
there is achieved a compromise between keeping the classical
uncertainty low and keeping quantum fluctuations also small.

\subsection{Measuring the gravitational field}
 This discussion
bears a close resemblance with the case of quantum gravity
concerning the existence of a minimum length, where there exists
a balance between the Heisenberg contribution $1/\Delta p$ to
the uncertainty in the position and the active response of
gravity to the presence of energy that produces an uncertainty
$\Delta x\gtrsim \ell_*^2\Delta p$. Actually, any measurement of
the gravitational field is not only extended in time, but also
extended in space. These measurements are made by determining
the change in the momentum of a test body of a given size. That
measurements of the gravitational field have to be extended in
spacetime, i.e., that they have to be continuous, is due to the
dynamical nature of this field. Before analyzing the
gravitational field, let us first briefly discuss the
electromagnetic field whose measurement can also be regarded as
continuous.

In the case of an electromagnetic field, the action has the form
$S=\int d^4x F^2$, where $F$ is the electromagnetic field
strength. Then, the action uncertainty principle $\Delta S
\gtrsim 1$ implies that $\Delta (F^2)\ l^4\gtrsim 1$, which can
be conveniently written as $\Delta F\ l^3\gtrsim q/(Flq)$, where
$l$ is the linear size of the test body and $q$ is its electric
charge. Here, we have already made the assumption that the
quantum fluctuations of the test body are negligible, i.e., that
its size $l$ is larger than its Compton wave length $1/m$, where
$m$ is its rest mass. $Flq$ is just the electromagnetic energy
of the test body. If we impose the condition that the
electromagnetic energy of the test body be smaller than its rest
mass $m$, the uncertainty relation above becomes in this case
$\Delta F\ l^3\gtrsim q/m$. The conditions $l\gtrsim 1/m$ and
$l\gtrsim q^2/m$ that we have imposed on the test body and that
can be summarized by saying that it must be classical from both
the quantum and the relativistic point of view are the
reflection of the following assumptions: the measurement of the
field averaged over a spacetime region, whose linear dimensions
and time duration are determined by $l$, is performed by
determining the initial and final momentum of a uniformly
charged test body; the time interval required for the momentum
measurement is small compared to $l$; any back-reaction can be
neglected if the mass of the test body is sufficiently high; and
finally, the borders of the test body are separated by a
spacelike interval.

Let us now consider \cite{me92} a measurement of the scalar
curvature averaged over a spacetime region of linear dimension
$l$, given by the resolution of the measuring device (the test
body). The action is $S=\ell_*^{-2}\int d^4x\sqrt{-g}R$, where
the integral is extended over the spacetime region under
consideration, so that it can be written as $S=\ell_*^{-2} R
l^4$, $R$ being now the average curvature. The action
uncertainty principle $\Delta S\gtrsim 1$ gives the uncertainty
relation for the curvature $\Delta R\ l^4\gtrsim \ell_*^2$,
which translates into the uncertainty relation $\Delta \Gamma\
l^3\gtrsim \ell_*^2$ for the connection $\Gamma$, or in terms of
the metric tensor, $\Delta g\ l^2\gtrsim \ell_*^2$. The left
hand side of this relation can be interpreted as the uncertainty
in the proper separation between the borders of the region that
we are measuring, so that it states the minimum position
uncertainty relation $\Delta x\gtrsim{\rm
min}(l,\ell_*^2/l)\gtrsim \ell_*$. It is worth noting that it is
the concurrence of the three fundamental constants of nature
$\hbar$, $c$ (which have already been set equal to 1), and
$\mbox{\small G}$ that leads to a resolution limit. If any of
them is dropped then this resolution limit disappears.

We see from the uncertainty relation for the electromagnetic
field that an infinite accuracy can be achieved if an
appropriate test body is used. This is not the case for the
gravitational interaction. Indeed, the role of $F$ is now played
by $\Gamma/\ell_*$, where $\Gamma$ is the connection, and the
role of $q$ is played by $\ell_* m$. It is worth noting
\cite{pr60} that by virtue of the equivalence principle, active
gravitational mass, passive gravitational mass and energy (rest
mass in the Newtonian limit) are all equal, and hence, for the
gravitational interaction, the ratio $q/m$ is the universal
constant $\ell_*$. The two requirements of Bohr and Rosenfeld
are now $l\gtrsim 1/m$ and $l\gtrsim \ell_*^2 m$ so that
$l\gtrsim \ell_*$. This means that the test body should not be a
black hole, i.e. its size should not exceed its gravitational
radius, and that both its mass and linear dimensions should be
larger than Planck's mass and length, respectively. As in the
electromagnetic case, Bohr and Rosenfeld requirements can be
simply stated as follows: the test body must behave classically
from the points of view of quantum mechanics, special relativity
and gravitation. Otherwise, the interactions between the test
body and the object under study would make this distinction (the
test body on the one hand and the system under study on the
other) unclear as happens in ordinary quantum mechanics: the
measurement device must be classical or it is useless as a
measuring apparatus. In this sense, within the context of
quantum gravity, Planck's scale establishes the border between
the measuring device and the system that is being measured.

We can see that the problem of measuring the gravitational
field, i.e., the structure of spacetime, can be traced back to
the fact that any such measurement is nonlocal, i.e. the
measurement device is aware of what is happening at different
points of spacetime and takes them into account. In other words,
the measurement device averages over a spacetime region. The
equivalence principle also plays a fundamental role: the
measurement device cannot decouple from the measured system and
back reaction is unavoidable.

\subsection{Vacuum fluctuations}
 One should expect not only
fluctuations of the gravitational field owing to the quantum
nature of other fields and measuring devices but also owing to
the quantum features of the gravitational field itself. As
happens for any other field, in quantum gravity there will exist
vacuum fluctuations that provide another piece of uncertainty to
the gravitational field strength. It can also be computed by means of
the action uncertainty principle. Indeed, in the above analyses,
we have only considered first order terms in the uncertainty
because it was assumed that there was a nonvanishing classical
field that we wanted to measure. However, in the case of vacuum,
the field vanishes and higher order terms are necessary. Let us
discuss this issue for the electromagnetic case first. The
uncertainty in the action can be calculated as $\Delta
S=S[F+\Delta F]-S[F]$, so that $\Delta S=\int d^4x [2 F\Delta
F+(\Delta F)^2]$. The action uncertainty principle then yields
the relation $\Delta F\ l^2 \gtrsim -Fl^2+\sqrt{(Fl^2)^2+1}$. In
the already studied limit of large electromagnetic field (or
very large regions) $Fl^2\gg 1$, the uncertainty relation for
the field becomes $\Delta F\ l^3\gtrsim 1/(Fl)\gtrsim q/m$
obtained above. On the other hand, the limit of vanishing
electromagnetic field (or very small regions of observation)
$Fl^2\ll1$ provides the vacuum fluctuations of the
electromagnetic field $\Delta F\ l^2\gtrsim 1$.

In the gravitational case, the situation is similar. The
gravitational action can be qualitatively written in terms of the
connection $\Gamma$ as $S=\ell_*^{-2}\int d^4x(\partial \Gamma+\Gamma^2)$
so that the uncertainty in the action has the form
\begin{equation}
\Delta S
\sim \ell_*^{-2}[\Delta \Gamma\ l^3+\Gamma\Delta \Gamma\ l^4
+ (\Delta \Gamma)^2l^4]\,.
\end{equation}
It is easy to argue that $\Gamma l$ must be at most of order 1
so that the contribution of the second term is qualitatively
equivalent to that of the first one. Indeed, $\Gamma l$
is the gravitational potential which
is given by $\Gamma l =\Gamma_{\rm ext}l(1-\Gamma
l)$, $\Gamma_{\rm ext}$ being the external gravitational field.
The last term is just an expression of the equivalence
principle, according to which, any kind of energy, including the
gravitational one, also generates a gravitational field. Thus,
$\Gamma l=\Gamma_{\rm ext}l/(1+\Gamma_{\rm
ext}l)$ which is always smaller than one. The action uncertainty
principle then implies that $\Delta \Gamma\ l^2 \gtrsim -l
+\sqrt{l^2+\ell_*^2}$ and that, in terms of the metric tensor,
\begin{equation}
\Delta g \gtrsim -1 +\sqrt{1+\ell_*^2/l^2}\,.
\end{equation}
For test bodies much larger than Planck size, i.e., for $l\gg
\ell_*$, this uncertainty relation becomes the already obtained
$\Delta g\gtrsim \ell_*^2/l^2$, valid for classical test bodies.
However, for spacetime regions of very small size --- closed to
Planck length $l\gtrsim \ell_*$ --- this uncertainty relation
acquires the form $\Delta g\gtrsim \ell_*/l$. This uncertainty
in the gravitational field comes from the vacuum fluctuations of
spacetime itself and not from the disturbances introduced by
measuring devices with $l\gg \ell_*$
\cite{mt73,wh55,wh57,wh62,wh64,wh68}. For alternative
derivations of this uncertainty relation see, e.g., Refs.
\cite{mt73,visser96}.

We then see that proper distances have an uncertainty
$\sqrt{\Delta g l^2}$ that approaches Planck length for very
small (Planck scale) separations thus suggesting that Planck
length represents a lower bound to any distance measurement. At
the Planck scale, the gravitational field uncertainty is of
order 1, i.e., the fluctuations are as large as the geometry
itself. This is indicating that the low-energy theory that we
have been using breaks down at the Planck scale and that a full
theory of quantum gravity is necessary to study such regime.


\section{Spacetime foam}
\indent

In his work ``On the hypotheses which lie at the basis of the
geometry'' \cite{riemann73}, written more than a century ago,
Riemann already noticed that ``[\ldots]. If this independence of
bodies from position does not exist, we cannot draw conclusions
from metric relations of the great, to those of the infinitely
small; in that case the curvature at each point may have an
arbitrary value in three directions, provided that the total
curvature of every measurable portion of space does not differ
sensibly from zero. Still more complicated relations may exist
if we no longer suppose the linear element expressible as the
square root of a quadratic differential. Now it seems that the
empirical notions on which the metrical determinations of space
are founded, the notion of a solid body and of a ray of light,
cease to be valid for the infinitely small. We are therefore
quite at liberty to suppose that the metric relations of space
in the infinitely small do not conform to the hypotheses of
geometry; and we ought in fact to suppose it, if we can thereby
obtain a simpler explanation of phenomena.''

In the middle of this century, Weyl \cite{we49} took these ideas
a bit further and envisaged (multiply connected) topological
structures of ever-increasing complexity as possible
constituents of the physical description of surfaces. He wrote
in this respect \cite{we49} ``A more detailed scrutiny of a
surface might disclose that what we had considered an elementary
piece in reality has tiny handles attached to it which change
the connectivity character of the piece, and that a microscope
of ever greater magnification would reveal ever new topological
complications of this type, {\it ad infinitum}.''

Few years later, Wheeler described this topological complexity
of spacetime at small length scales as the foamlike structure of
spacetime \cite{wh57}. According to Wheeler
\cite{mt73,wh55,wh57,wh62,wh64,mw57}, at the Planck scale, the
fluctuations of the geometry are so large and involve so large
energy densities that gravitational collapse should be
continuously being done and undone at that scale. Because of
this perpetuity and ubiquity of Planck scale gravitational
collapse, it should dominate Planck scale physics. In this
continuously changing scenario, there is no reason to believe
that spacetime topology remains fixed and predetermined. Rather,
it seems natural to accept that the topology of spacetime is
also subject to quantum fluctuations that change all its
properties. Therefore, this scenario, in which spacetime is
endowed with a foamlike structure at the Planck scale, seems to
be a natural ingredient of the yet-to-be-built quantum theory of
gravity. Furthermore, from the functional integration point of
view \cite{mi57}, in quantum gravity all histories contribute
and, among them, there seems unnatural not to consider
nontrivial topologies as one considers not trivial geometries
\cite{wh55,wh57,misner60} (see, however, Ref. \cite{dewitt84}).
On the other hand, it has been shown \cite{horowitz90} that
there exit solutions to the equations of general relativity on
manifolds that present topology changes. In these solutions, the
metric is degenerate on a set of measure zero but the curvature
remains finite. This means that allowing degenerate metrics
amounts to open a doorway to classical topology change.
Furthermore, despite the difficulties of finding an appropriate
interpretation for these degenerate metrics in the classical
Lorentzian theory, they will naturally enter the path integral
formulation of quantum gravity. This is therefore an indication
that topology change should be taken into account in any quantum
theory of gravity \cite{horowitz90} (for an alternative
description of topology change within the framework of
noncommutative geometry, see Ref. \cite{madore98,mangano98}).

Adopting a picture in which spacetime topology depends on the
scale on which one performs the observations, we would conclude
that there would be a trivial topology on large length scales
but more and more complicated topologies as we approach the
Planck scale.

Spacetime foam may have important effects in low-energy physics.
Indeed, the complicated topological structure may provide
mechanisms for explaining the vanishing of the cosmological
constant \cite{coleman88b,1ca97,2ca97} and for fixing all the
constants of nature \cite{coleman88b,ha90} (for a recent
proposal for deriving the electroweak coupling constant from
spacetime foam see Ref. \cite{rosales99}). Spacetime foam may
also induce loss of quantum coherence \cite{ha82} and may well
imply the existence of an additional source of uncertainty.
Related to this, it might produce frequency-dependent energy
shifts \cite{1ga98,2ga98,3ga98} that would slightly alter the
dispersion relations for the different low-energy fields.
Finally, spacetime foam has been proposed as a mechanism for
regulating both the ultraviolet \cite{crane86} (see also Refs.
\cite{de57,le83,pa85,pa88}) and the infrared \cite{magnon88}
behavior of quantum field theory.

It is well-known that it is not possible to classify all
four-dimensional topologies \cite{ma58,ha78} and, consequently,
all the possible components of spacetime foam.
With the purpose of exemplifying the richness and complexity
of the vacuum of quantum gravity, in what follows, we will
briefly discuss a few different kinds of fluctuations encompassed by
spacetime foam, where the
word fluctuations will just denote spacetime configurations that
contribute most to the gravitational path integral \cite{wh57}:
simply connected nontrivial topologies, multiply connected
topologies with trivial second homology group (i.e. with
vanishing second Betti number),  spacetimes with a
nontrivial causal structure, i.e. with closed timelike curves,
in a bounded region, and, finally, nonorientable tunnels.

Hawking \cite{ha78} argued that the dominant contribution to the
quantum gravitational path integral over metrics and topologies
should come from topologies whose Euler characteristic
$\chi_{\mathsc E}$ was approximately given by the spacetime
volume in Planck units, i.e., from topologies with
$\chi_{\mathsc E}\sim (l/\ell_*)^4$. In this analysis, he
restricted to compact simply-connected manifolds with negative
cosmological constant $\lambda$. The choice of compact manifolds
obeys to a normalization condition similar to introducing a box
of finite volume in nongravitational physics. The cosmological
constant is introduced for this purpose and it being negative is
because saddle-point Euclidean metrics with high Euler
characteristic and positive $\lambda$ do not seem to exist, so
that positive-$\lambda$ configurations will not contribute
significantly to the Euclidean path integral. Finally, simple
connectedness can be justified by noting that multiply-connected
compact manifolds can be unwrapped by going to the universal
covering manifold that, although will be noncompact, can be made
compact with little cost in the action. He then concluded that,
among these manifolds, the dominant topology is $S^2\times S^2$
\cite{ha96} which has an associated second Betti number
$B_2=\chi_{\mathsc E}-2=2$. These results are based on the
semiclassical approximation and, as such, should be treated with
some caution.

Compact simply-connected bubbles with the topology $S^2\times
S^2$ can be interpreted as closed loops of virtual black holes
\cite{ha96} if one realizes \cite{gi86} that the process of
creation of a pair of real charged black holes accelerating away
from each other in a spacetime which is asymptotic to $\Re^4$ is
provided by the Ernst solution \cite{er76}. This solution has
the topology $S^2\times S^2$ minus a point (which is sent to
infinity) and this topology is the topological sum of the bubble
$S^2\times S^2$ plus $\Re^4$. Virtual black holes will not obey
classical equations of motion but will appear as quantum
fluctuations of spacetime and thus will become part of the
spacetime foam. As a consequence, one can conclude that the
dominant contribution to the path integral over compact
simply-connected topologies would be given by a gas of virtual
black holes with a density of the order of one virtual black
hole per Planck volume.

A similar analysis within the context of quantum conformal
gravity has been performed by Strominger \cite{st84} with the
conclusion that the quantum gravitational vacuum indeed has a
very involved structure at the Planck scale, with a
proliferation of nontrivial compact topologies.

Carlip \cite{1ca97,2ca97} has studied the influence of the
cosmological constant $\lambda$ on the sum over topologies. It
should be stressed that this cosmological constant is not
related to the observed cosmological constant \cite{ha78}.
Rather, it is introduced as a source term of the form
$\ell_*^{-2}\lambda V$, where $V$ is the spacetime volume, added
to the vacuum gravitational action $\ell_*^{-2}\int R\sqrt g$.
In the semiclassical approximation, this sum is dominated by the
saddle-points, which are Einstein metrics. The classical
Euclidean action for these metrics has the form $\tilde
v/(\ell_*^2\lambda)$, where, up to irrelevant numerical factors,
$\tilde v=\lambda^2V$ is the normalized spacetime volume of the
manifold and is independent of $\lambda$. In fact, $\tilde v$
characterizes the topology of the manifold. For instance, for
hyperbolic manifolds it can be identified with the Euler
characteristic. Carlip has shown that, in the semiclassical
approximation, the behavior of the density of topologies, which
counts the number of manifolds with a given value for $\tilde
v$, crucially depends on the sign of the cosmological constant.

For negative values of $\lambda$, the partition function
receives relevant contributions from spacetimes with arbitrarily
complicated topology, so that processes that could be expected
to contribute to the vacuum energy might produce more and more
complicated spacetime topologies, as we briefly discuss in what
follows, thus providing a mechanism for the vanishing of the
cosmological constant. The Euclidean path integral in the
semiclassical approximation can be written as
\begin{equation}
Z[\lambda] =\sum_{\tilde{v}}\rho(\tilde{v}) e^{\tilde{v}
/(\ell_*^{2}\lambda)}\,,
\end{equation}
where $\rho(\tilde{v})$ is a density of topologies. It can be
argued that for negative $\lambda$, the density of topologies
$\rho(\tilde{v})$ grows with the topological complexity $\tilde
v$ at least as $\rho(\tilde{v})\gtrsim
\exp(\tilde{v}\ln \tilde{v})$, i.e., it is
superexponential \cite{1ca97,2ca97}. Then, after introducing an
infrared cutoff to ensure the convergence of the sum above, the
topologies that will contribute most to $Z[\lambda]$ will lie
around some maximum value of the topological complexity
$\tilde{v}_{\rm max}$. The true cosmological constant $\Lambda$,
obtained from the microcanonical ensemble, is in this case
\begin{equation}
-\frac{1}{\Lambda
\ell_*^2}=\left.\frac{\partial\ln\rho(\tilde{v})}
{\partial\tilde{v}}\right|_{\tilde{v}_{\rm max}}\gtrsim
1+\ln\tilde{v}_{\rm max}
\end{equation}
and the ``topological capacity''
\begin{equation}
c_V=-\frac{1}{\Lambda^2
\ell_*^4}\left.\left(\frac{\partial^2\ln\rho(\tilde{v})}
{\partial\tilde{v}^2}\right)^{-1}\right|_{\tilde{v}_{\rm
max}}=-\ell_*^{-2}\left(
\frac{\partial\Lambda}{\partial\tilde v_{\rm max}}\right)^{-1}
\lesssim -\tilde{v}_{\rm max}(1+\ln\tilde{v}_{\rm max})\,,
\end{equation}
where these quantities have been defined by analogy with the
thermodynamical temperature and heat capacity, respectively. In
this analogy, $-\Lambda$ plays the role of temperature while the
topological complexity $\tilde v$ is analogous to the energy.
According to this picture, the behavior of spacetime foam would
be analogous to a thermodynamical system with negative heat
capacity, in which, as we put energy into the system, a greater
and greater proportion of it is employed in the exponential
production of new states rather than in increasing the energy of
already existing states. Similarly, since the topological
capacity is negative, which is a consequence of the
superexponential density of topologies, the microcanonical
cosmological constant will approach a vanishing value as the
maximum topological complexity $\tilde{v}_{\rm max}$ approaches
infinity. We then see that this process, which could be expected
to increase the vacuum energy $|\Lambda|$, actually contributes
to decrease it, until it approaches the smallest value
$|\Lambda|=0$. The case of positive $\lambda$ presents a
different behavior. The topological complexity has a finite
maximum value, namely, that of the four-sphere $\tilde v_{\rm
max}=\chi_{\mathsc E}^{\rm max}=2$ and the density of topologies
$\rho(\tilde{v})$ increases as $\tilde{v}$ decreases.

The superexponential lower bound to the density of topologies
given above receives the main contribution from multiply
connected manifolds, among which, Euclidean wormholes
\cite{ha90b,ha90c} have deserved much attention during the last
decade (see, e.g., Ref. \cite{barcelo98}). Wormholes are four-dimensional
spacetime handles that have vanishing second Betti number, while
the first Betti number provides the number of handles. They
were regarded as a possible mechanism for the complete
evaporation of black holes
\cite{ha87,polchinski94,gonzalez91c,cavaglia96}. An evaporating
black hole would have a wormhole attached to it and this
wormhole would transport the information that had fallen into
the black hole to another, quite possibly far away, region of
spacetime. More recently, Hawking \cite{ha96} has proposed an
alternative scenario in which black holes, at the end of their
evaporation process, will have a very small size and will
eventually dilute in the sea of virtual black holes that form
part of spacetime foam. Wormholes also constitute the main
ingredient in Coleman's proposal for explaining the vanishing of
the cosmological constant and for fixing all the constants of
nature \cite{coleman88b,ha90} (see also Ref. \cite{unruh89}).
Wormholes have been studied in the so-called dilute gas
approximation in which wormhole ends are far apart form each
other. It should be noted, however, that, although the semiclassical
approximation probably ceases to be valid at the Planck scale,
it gives a clear indication that one should expect a topological
density of one wormhole per unit four-volume, i.e., the first
Betti number $B_1$ should be approximately equal to the
spacetime volume $B_1\sim V$ at the Planck scale. Multiply
connected topology fluctuations may suffer instabilities
against uncontrolled growth both in Euclidean quantum gravity
\cite{klebanov89,fischler89,polchinski89} (see however Ref.
\cite{coleman89}) and in the Lorentzian sector
\cite{redmount93,redmount94}. These instabilities might put
serious limitations to the kind of multiply connected topologies
encompassed by spacetime foam.

One should also expect other configurations with nontrivial
causal structure to contribute to spacetime foam. For instance,
quantum time machines \cite{go97,go98}, have been recently
proposed as possible components of spacetime foam. From the
semiclassical point of view, most of the hitherto proposed time
machines \cite{1mt88,2mt88} are unstable because quantum vacuum
fluctuations generate divergences in the stress-energy tensor,
i.e., are subject to the chronology protection conjecture
\cite{ha92,cassidy98} (for a beautiful and detailed report on
time machines see Ref. \cite{visser96}). However, quantum time
machines \cite{go97,go98} confined to small spacetime regions,
for which the chronology protection conjecture does not apply
\cite{lg98}, are likely to occur within the realm of spacetime
foam, where strong causality violations or even the absence of a
causal structure are expected. We have in fact argued that the
spacetime metric undergoes quantum fluctuations of order 1 at
the Planck scale. Since the slope of the light cone is
determined by the speed of light obtained from
$ds^2=g_{\mu\nu}dx^\mu dx^\nu=0$, the uncertainty in the metric
will also introduce an uncertainty in the slope of the light
cone of order 1 at the Planck scale so that the notion of
causality is completely lost.

As happens with the causal structure, orientability is likely to
be lost at the Planck scale \cite{friedman88,gonzalez98}, where
the lack of an arbitrarily high resolution would blur the
distinction between the two sides of any surface. Therefore,
nonorientable topologies can be regarded as additional
configurations that may well be present in spacetime foam and
thus contribute to the vacuum structure of quantum gravity.
Indeed, quantum mechanically stable nonorientable spacetime
tunnels that connect two asymptotically flat regions with the
topology of a Klein bottle can be constructed \cite{gonzalez98}
as a generalization of modified Misner space \cite{go97,go98}.

The presence of quantum time machines or nonorientable tunnels
in spacetime amounts to the existence of Planck-size regions in
which violations of the weak energy condition occur. Although
from the classical point of view, the weak energy condition
seems to be preserved, it is well-known (see, e.g., Ref.
\cite{visser96}) that quantum effects may well involve such
exotic types of energy.


\section{Loss of quantum coherence}
\indent

The quantum structure of spacetime would be relevant at energies
close to Planck scale and one could expect that the quantum
gravitational virtual processes that constitute the spacetime
foam could not be described without knowing the details of the
theory of quantum gravity. However, the gravitational nature of
spacetime fluctuations provides a mechanism for studying the
effects of these virtual processes in the low-energy physics.
Indeed, virtual gravitational collapse and topology change would
forbid a proper definition of time at the Planck scale. More
explicitly, in the presence of horizons, closed timelike curves,
topology changes, etc., any Hamiltonian vector field that
represents time evolution outside the fluctuation would vanish
at points inside the fluctuation. This means that it would not
be possible to describe the evolution by means of a Hamiltonian
unitary flow from an initial to a final state and, consequently,
quantum coherence would be lost. These effects and their order
of magnitude would not depend on the detailed structure of the
fluctuations but rather on their existence and global
properties. In general, the regions in which the asymptotically
timelike Hamiltonian vector fields vanish are associated with
infinite redshift surfaces and, consequently, these small
spacetime regions would behave as magnifiers of Planck length
scales transforming them into low-energy modes as seen from
outside the fluctuations \cite{1pa98,2pa98}. Therefore,
spacetime foam and the related lower bound to spacetime
uncertainties would leave their imprint, which may be not too
small, in low-energy physics and low-energy experiments would
effectively suffer a nonvanishing uncertainty coming from this
lack of resolution in spacetime measurements. In this situation,
loss of quantum coherence would be almost unavoidable
\cite{ha82}.

The idea that the quantum gravitational fluctuations contained
in spacetime foam could lead to a loss of quantum coherence was
put forward by Hawking and collaborators \cite{ha82,hp79,hp80}.
This proposal was based in part on the thermal character of the
emission predicted for evaporating black holes
\cite{hawking75,wald75,hawking76}. If loss of coherence occurs
in macroscopic black holes, it seems reasonable to conclude that
the small black holes that are continuously being created and
annihilated everywhere within spacetime foam will also induce
loss of quantum coherence \cite{ha82,hawking76}. On the other
hand, scattering amplitudes of low-energy fields by
topologically nontrivial configurations ($S^2\times S^2$, $K^3$
and $CP^2$ bubbles) lead to the conclusion that pure states turn
into a partly incoherent mixture upon evolution in these
nontrivial backgrounds under certain simplifying assumptions
and, consequently, that quantum coherence is lost.

They made explicit calculations for specific asymptotically flat
spacetimes with nontrivial simply-connected topologies
\cite{ha82,ha96,hp79,hp80,hawking84,warner82,hr97} or causal
structure \cite{ha95} which showed that it was not possible to
separate the complex-time graphs for the obtained Lorentzian
Green functions into two disconnected parts. More explicitly,
the Euclidean Green functions obtained in these backgrounds mix
positive and negative frequencies when the analytic continuation
to Lorentzian signature is performed, since the Green functions
develop extra acausal singularities. This situation is analogous
to that in black hole physics where Lorentzian Green functions
show periodic poles in imaginary time \cite{hp80}. Although
these calculations were performed in a finite dimensional
approximation to metrics of given topology, the contributions of
these extra singularities can be determined by dimensional
analysis and therefore they seem to be characteristic of each
topology and hold for any metric in them \cite{hp80}. In
contrast, Gross \cite{gross84} calculated scattering amplitudes
in specific four-dimensional solutions that can be interpreted
as three-dimensional Kaluza-Klein instantons and concluded that
there was no loss of quantum coherence in such models. Hawking
\cite{hawking84} in turn replied to this criticism that the
solutions used by Gross were special cases in the sense that the
associated three-dimensional Kaluza-Klein instantons were flat
and therefore topologically trivial. He further argued, with
examples, that solutions with topologically nontrivial
three-dimensional instantons can be constructed and that lead to
a nonunitary evolution.

\subsection{Superscattering operator}
 Let us consider a scattering
process in an asymptotically flat spacetime with nontrivial
topology. If we denote the density matrices at the far past and
far future by $\rho_-$ and $\rho_+$, respectively, there will be
a superscattering operator $\sso $ that relates both of them
$\rho_+=\sso\cdot\rho_-$, i.e., that provides the evolution
between the two asymptotically flat regions across the
nontrivial topology fluctuation \cite{ha82}. Let $|0_\pm\rangle$
represent the vacuum at each region and $\{ |A_\pm\rangle\}$ a
basis of the Fock space, so that we can write
$|A_\pm\rangle=\Upsilon_{\pm A}^\dag |0_\pm\rangle$, where
$\Upsilon^A$ is a string of annihilation operators and,
consequently, $\Upsilon^\dag_A$ is a string of creation
operators. The density matrices $\rho_\pm$ can then be written
as
\begin{equation}
\rho_\pm=\sum_{AB}\rho_{\pm\;B}^{\;A}|A_\pm\rangle\langle B_\pm|=
\rho_{\pm\;B}^{\;A}\Upsilon_{\pm A}^\dag|0_\pm\rangle\langle
0_\pm|\Upsilon^B_\pm\,,
\end{equation}
where a sum over repeated indices is assumed.

The density matrices at both asymptotic regions can then be
related by noting that the density matrix at the far future
$\rho_+$ is given by the expectation values in the far-past
state $\rho_-$ of a complete set of future operators built out
of creation and annihilation operators, namely,
\begin{equation}
\rho_{+\;D}^{\;C}={\rm tr}(\Upsilon_{+D}^\dag \Upsilon^C_+\rho_-)=
\rho_{-\;B}^{\;A}\langle 0_-|\Upsilon^B_-\Upsilon_{+D}^\dag
\Upsilon^C_+\Upsilon_{-A}^\dag|0_-\rangle\,.
\end{equation}
Therefore, the superscattering matrix
$\sso^{C\;\;\;\;\;B}_{\;\;DA}\equiv
\langle 0_-|\Upsilon^B_-\Upsilon_{+D}^\dag \Upsilon^C_+
\Upsilon_{-A}^\dag|0_-\rangle$, relates the density
matrices in both asymptotic regions, i.e., $\rho_{+\;D}^{\;C}=
\sso^{C\;\;\;\;\;B}_{\;\;DA} \rho_{-\;B}^{\;A}$.
Note that the superscattering matrix
$\sso^{C\;\;\;\;\;B}_{\;\;DA}$ is Hermitian in both pairs of
indices $CD$ and $AB$ to ensure that the Hermiticity of the
density matrix is preserved. Also, the conservation of
probability, i.e., ${\rm tr}(\rho_\pm)=1$, implies that
$\sso^{C\;\;\;\;\;B}_{\;\;CA}=\delta_A^{\;\;B}$.

The relation between this superscattering operator and the Green
functions discussed above is easily obtained if we write the
annihilation operators $a_\pm(k)$ that form $\Upsilon^A_\pm$ at
each asymptotic region in terms of the corresponding field
operators. For instance, in the case of a complex scalar field,
this expression (up to numerical normalization factors) has the
well-known form
\begin{equation}
a_\pm(k)=-i\int_{\Sigma_\pm} d\Sigma^\mu(x)
e^{-kx}\stackrel{\leftrightarrow}{\nabla}_\mu\phi(x)\,,
\end{equation}
where $\Sigma_\pm$ represent spacelike surfaces in the infinite
past and future.

We now introduce the identity operator $1=\sum_n
|n\rangle\langle n|$, with $|n\rangle$ being energy eigenstates,
in the expression for $\sso$
\begin{equation}
\sso^{C\;\;\;\;\;B}_{\;\;DA}= \sum_n \langle
0_-|\Upsilon^B_-\Upsilon_{+D}^\dag
|n\rangle\langle n|\Upsilon^C_+
\Upsilon_{-A}^\dag|0_-\rangle
\end{equation}
and note that the only state that can contribute is that with
zero energy, i.e., $n=0$ for energy to be conserved. If
spacetime is globally hyperbolic, so that asymptotic
completeness holds, there is a one-to-one map between states at
any spacetime region, in particular, between the vacua
$|0\rangle$ and $|0_+\rangle$. Therefore, the only contribution
from $1=\sum_n |n\rangle\langle n|$ can be regarded as coming
from $|0_+\rangle\langle 0_+|$:
\begin{equation}
\sso^{C\;\;\;\;\;B}_{\;\;DA}= \langle
0_-|\Upsilon^B_-\Upsilon_{+D}^\dag
|0_+\rangle\langle 0_+|\Upsilon^C_+
\Upsilon_{-A}^\dag|0_-\rangle\,.
\end{equation}
In this case, the superscattering operator factorizes into two
unitary factors:
\begin{equation}
\sso^{C\;\;\;\;\;B}_{\;\;DA}=S^C_{\;\;A} S^{*\;\;B}_{\;D}\,,
\end{equation}
with $S^C_{\;\;A}=\langle 0_+|\Upsilon^C_+
\Upsilon_{-A}^\dag|0_-\rangle =\langle
C_+|A_-\rangle$. Note that the scattering matrix $S$ is indeed
unitary, i.e., $S^C_{\;\;A} S^{*\;\;B}_{\;C}=
\sum_C\langle B_-|C_+\rangle\langle C_+|A_-\rangle
= \delta_A^{\;\;B}$ by virtue of the condition of conservation
of probability. The factorizability of the superscattering
operator $\sso$ always implies unitary evolution for the density
matrix. Indeed, if the superscattering operator can be
factorized as $\sso\cdot\rho=S\rho S^\dag$ for some scattering
operator $S$, then conservation of probability, which amounts to
require that ${\rm tr}(\sso\cdot\rho)=1$ provided that ${\rm
tr}(\rho)=1$, implies that
\begin{equation}
1={\rm tr}(\sso\cdot\rho)={\rm tr}(S\rho S^\dag)={\rm tr}(\rho
S^\dag S)
\end{equation}
and therefore $S^\dag S=1$, i.e., the scattering operator $S$ is
unitary. In this case, the operator $\sso$ also implies a
unitary evolution for the density matrix since it preserves
${\rm tr}(\rho^2)$:
\begin{equation}
{\rm tr} (\rho_+^2)={\rm tr}[(\sso \cdot\rho_-)
(\sso\cdot\rho_-)]={\rm tr} (S\rho_-S^\dag S\rho_- S^\dag)= {\rm
tr}(S\rho_-^2 S^\dag)={\rm tr}(\rho_-^2)\,.
\end{equation}

If, on the other hand, we cannot guarantee that states at
different spacetime regions are one-to-one related, then the
zero energy state $|0\rangle$ will not correspond in general to
the zero energy state $|0_+\rangle$ and the superscattering
operator will not admit a factorized form: $\sso\cdot\rho\neq
S\rho S^\dag$. When the superscattering operator does not
satisfy the factorization condition, the evolution does not
preserve ${\rm tr}(\rho^2)$ in general and quantum coherence is
lost. This can be seen explicitly in the analysis below.

\subsection{Quasilocal superscattering}
Let us assume that the dynamics that underlies a superscattering
operator $\sso$ is quasilocal. By quasilocal we mean that any
possible effect leading to a nonfactorizable superscattering
operator is confined to a spacetime region whose size $r$ is
much smaller than the characteristic spacetime size $l$ of the
low-energy fields, i.e., we will assume that $r/l\ll 1$. Then,
the superscattering equation $\rho_+=\sso \cdot
\rho_-$ can be obtained by integrating a differential equation
of the form $\dot\rho(t)=L(t)\cdot
\rho(t)$, where $L(t)$ is a linear operator \cite{el84}.
Furthermore, it can be shown that $L(t)$ can be generally
written as \cite{bs84}
\begin{eqnarray}
L\cdot\rho\!\!\!&=&\!\!\!-i\big[ H_0,\rho\big]-\frac{1}{2}
h_{\alpha\beta} (Q^\beta Q^\alpha\rho+\rho Q^\beta Q^\alpha-2
Q^\alpha\rho Q^\beta)
\nonumber\\
\!\!\!&=&\!\!\!-i\big[ H_0,\rho\big]-\frac{i}{2}{\rm Im
}(h_{\alpha\beta})\big[Q^\alpha,\big[Q^\beta, \rho\big]_+\big]
-\frac{1}{2}{\rm
Re}(h_{\alpha\beta})\big[Q^\alpha,\big[Q^\beta,\rho\big]\big]\,,
\end{eqnarray}
where $H_0$ and $Q^\alpha$ form a complete set of Hermitian
matrices, $Q^\alpha$ have been chosen to be orthogonal, i.e.,
${\rm tr}(Q^\alpha Q^\beta)=\delta^{\alpha\beta}$, and
$h_{\alpha\beta}$ is a Hermitian matrix. A sufficient, but not
necessary, condition for having a decreasing value of ${\rm
tr}(\rho^2)$ and, consequently, loss of coherence is that
$h_{\alpha\beta}$ be real and positive. As a simple example, we
can consider the case in which we have only one operator $Q$.
Then,
\begin{equation}
\frac{d}{dt}{\rm tr}(\rho^2)=-{\rm tr}(\rho^2Q^2-\rho Q\rho Q)\,.
\end{equation}
If we diagonalize the density matrix and call $\{|i\rangle\}$ to
the preferred basis in which $\rho$ is diagonal, so that
$\rho=\sum_i p_i|i\rangle\langle i|$, this equation becomes
\begin{equation}
\frac{d}{dt}{\rm tr}(\rho^2)=-\sum_{ij} p_i |Q_{ij}|^2(p_i-p_j)
= -\sum_{i>j}|Q_{ij}|^2(p_i-p_j)^2\,,
\end{equation}
where $Q_{ij}=\langle i|Q|j\rangle$. We then see that provided
that $Q$ is not diagonal in the basis $\{|i\rangle\}$,
$\frac{d}{dt}{\rm tr}(\rho^2)<0$, except for very specific
states, such as the obvious $p_i=p_j$, which has maximum
entropy.

There has been an interesting debate on the possible violations
of energy and momentum conservation or locality in processes
that do not lead to a factorizable $\sso$ matrix. According to
Gross \cite{gross84} and Ellis {\it et al.} \cite{el84}, a
nonfactorizable $\sso$ matrix allows for continuous symmetries
whose associated generators are not conserved. In other words,
``invariance principles are no longer equivalent to conservation
laws'' \cite{el84}.

Let us illustrate this issue with the simple example \cite{el84}
of two spin-1/2 particles in a state described by the density
matrix $\rho_-=\frac{1}{4}(1-\vec s_1\vec s_2)$, where $\vec
s_{1,2}$ are the spin vectors of the particles 1 and 2,
respectively. This density matrix represents a pure state since
${\rm tr}(\rho_-^2)=1$. In fact, the two particles are in a
rotationally invariant pure state with vanishing total spin.
Assume that the final state can be obtained by a superscattering
operator $\sso$. Then, $\rho_+=\sso\cdot\rho_-$ must have the
form $\rho_+=\frac{1}{4}(1-\beta\vec s_1\vec s_2)$, for it to
conserve probability ${\rm tr}(\rho_+)=$1 and be rotationally
invariant. Furthermore, since ${\rm tr}(\rho_+^2)=
(1+3\beta^2)/4\leq1$, we must have $\beta\leq 1$, the equality
holding only when $\rho_+$ is a pure state. The initial state is
such that ${\rm tr}[(\vec s_1+ \vec s_2)^2\rho_-]=-1$, which
means that, in any given direction, there is initially a perfect
anticorrelation between the spin of the two particles, so that
the total spin vanishes. However, for the final state, ${\rm
tr}[(\vec s_1+\vec s_2)^2\rho_+]=1-\beta$. We then see that,
despite the rotational invariance of the states and the
evolution, we will not obtain total anticorrelation in the final
state and, hence, spin conservation, unless $\beta=1$, i.e.,
unless quantum coherence is preserved.

In particular, these authors \cite{gross84,el84} argued that
energy and momentum conservation does not follow from Poincar\'{e}
invariance. However, energy and momentum conservation is a
consequence of the field equations in the asymptotic regions
\cite{hawking84}. This issue also arises when the evolution of
the density matrix is obtained by a differential equation whose
integral leads to a nonfactorizable $\sso$ operator. If this
equation is assumed to be local on scales a bit larger than
Planck length, then there appears a conflict between this
pretended locality on the one hand and energy and momentum
conservation on the other \cite{bs84}. This violation of energy
and momentum conservation comes from the high-energy modes,
whose characteristic evolution times is of the same order as the
size of the nontrivial topology region. Again, the existence of
asymptotic regions would enforce this conservation and this can
be effectively achieved if the propagating fields are regarded
as low-energy ones and, therefore, with characteristic size $l$
much larger the size $r$ of the fluctuation. Furthermore, Unruh
and Wald \cite{uw95} analyzed simple non-Markovian toy models
that lose quantum coherence and argued that conservation of
energy and momentum need not be in conflict with causality and
locality, in contrast with the claims of Ref. \cite{bs84} (see
also Ref. \cite{srednicki93,liu93}). Therefore, these topology
fluctuations can be regarded as nonlocal in the length scale
$r$, since, within this scale, the unitary $S$-matrix diagrams
will be mixed (thus leading to a nonfactorizable $\sso$ matrix),
while from the low-energy point of view, the fluctuations are
confined in a very small region so that they can be described as
local effective interactions in a master differential equation
as above. This relation will be the subject of the next two
sections.


\section{Quantum bath}
\indent

Spacetime foam contains, according to the scenario above, highly
nontrivial topological or causal configurations, which will
introduce additional features in the description of the
evolution of low-energy fields as compared with topologically
trivial, globally hyperbolic manifolds. The analogy with fields
propagating in a finite-temperature environment is compelling.
Actually, despite the different conceptual and physical origin
of the fluctuations, we will see that the effects of these two
systems are not that different.

In order to build an effective theory that accounts for the
propagation of low-energy fields in a foamlike spacetime, we
will substitute the spacetime foam, in which we possibly have a
minimum length because the notion of distance is not valid at
such scale, by a fixed background with low-energy fields living
on it. We will perform a 3+1 foliation of the effective
spacetime that, for simplicity, will be regarded as flat, $t$
denoting the time parameter and $x$ the spatial coordinates. The
gravitational fluctuations and the minimum length present in the
original spacetime foam will be modeled by means of nonlocal
interactions that relate spacetime points that are sufficiently
close in the effective background, where a well-defined notion
of distance exists \cite{1ga98,2ga98,3ga98} (for related ideas
see also Refs. \cite{martin98,martin98b} and for a review on
stochastic gravity see Ref. \cite{hu99}). Furthermore, these
nonlocal interactions will be described in terms of local
interactions as follows. Let $\{h_i[\phi;t]\}$ be a basis of
local gauge-invariant interactions at the spacetime point
$(x,t)$ made out of factors of the form
$\ell_*^{2n(1+s)-4}\left[\phi(x,t)\right]^{2n}$, $\phi$ being
the low-energy field strength of spin $s$. As a notational
convention, each index $i$ implies a dependence on the spatial
position $x$ by default; whenever the index $i$ does not carry
an implicit spatial dependence, it will appear underlined
${\underline{i}}$. Also, any contraction of indices (except for
underlined ones) will entail an integral over spatial positions.

\subsection{Influence functional}
 The low-energy density
$\rho[\phi,\varphi;t]$ at the time $t$ in the field
representation can be generally related to the density matrix at
$t=0$
\begin{equation}
\rho[\phi,\varphi; t]=\int D\phi'  D\varphi'
\sso[\phi,\varphi;t|\phi',\varphi';0] \rho[\phi',\varphi';0]\,,
\end{equation}
which we will write in the compact form $\rho(t)=\sso(t)\cdot
\rho(0)$. Here $\sso(t) $ is the propagator for the density matrix
and $D\phi\equiv\prod_x \phi(x,t)$. This propagator has the form
\begin{equation}
\sso[\phi,\varphi;t|\phi',\varphi';0]=\int {\cal D}\phi {\cal
D}\varphi e^{i\{S_0[\phi;t]-S_0[\varphi;t]\}}{\cal
F}[\phi,\varphi;t]\,,
\end{equation}
where ${\cal F}[\phi,\varphi;t]$ is the so-called influence
functional \cite{fv63,fh65,ca83}, ${\cal D}\phi\equiv\prod_{x,s}
\phi(x,s)$ and these path integrals are
performed over paths $\phi(s)$, $\varphi(s)$ such that at the
end points match the values $\phi$, $\varphi$ at $t$ and
$\phi'$, $\varphi'$ at $s=0$. The influence functional ${\cal
F}[\phi,\varphi;t]$ contains all the information about the
interaction of the low-energy fields with spacetime foam. Let us
now introduce another functional ${\cal W}[\phi,\varphi;t]$ that
we will call influence action and such that ${\cal
F}[\phi,\varphi;t]=\exp{\cal W}[\phi,\varphi;t]$. If the
influence action ${\cal W}[\phi,\varphi;t]$ were equal to the
zero, then we would have unitary evolution provided by a
factorized superscattering matrix. However, ${\cal W}$ does not
vanish in the presence of gravitational fluctuations and, in
fact, the nonlocal effective interactions will be modeled by
terms in ${\cal W}$ that follow the pattern
\begin{equation}
\int  dt_1\cdots dt_N \upsilon^{i_1\cdots i_N}(t_1\ldots
t_N)h_{i_1}[\phi;t_1]\cdots h_{i_N}[\phi;t_N]\,.
\end{equation}
Here, $\upsilon^{i_1\cdots i_N}(t_1\ldots t_N)$ are
dimensionless complex functions that vanish for relative
spacetime distances larger than the length scale $r$ of the
gravitational fluctuations. If the gravitational fluctuations
are smooth in the sense that they only involve trivial
topologies or contain no horizons, the coefficients
$\upsilon^{i_1\cdots i_N}(t_1\ldots t_N)$ will be $N$-point
propagators which, as such, will have infinitely long tails and
the size of the gravitational fluctuations will be effectively
infinite. In other words, we would be dealing with a local
theory written in a nonstandard way. The gravitational origin of
these fluctuations eliminate these long tails because of the
presence of gravitational collapse and topology change. This
means that, for instance, virtual black holes \cite{ha96} will
appear and disappear and horizons will be present throughout. As
Padmanabhan \cite{1pa98,2pa98} has also argued, horizons induce
nonlocal interactions of finite range since the Planckian
degrees of freedom will be magnified by the horizon (because of
an infinite redshift factor) thus giving rise to low-energy
interactions as seen from outside the gravitational fluctuation.
Virtual black holes represent a kind of components of spacetime
foam that because of the horizons and their nontrivial topology
will induce nonlocal interactions but, most probably, other
fluctuations with complicated topology will warp spacetime in a
similar way and the same magnification process will also take
place.

The coefficients $\upsilon^{i_1\cdots i_N}(t_1\ldots t_N)$ can
depend only on relative positions and not on the location of the
gravitational fluctuation itself. The physical reason for this is
conservation of energy and momentum: the fluctuations do not
carry energy, momentum, or gauge charges. Thus, diffeomorphism
invariance is preserved, at least at low-energy scales. One
should not expect that at the Planck scale this invariance still
holds. However, this violation of energy-momentum conservation is
safely kept within Planck scale limits \cite{uw95}, where the
processes will no longer be Markovian.

Finally, the coefficients $\upsilon^{i_1\cdots i_N}(t_1\ldots
t_N)$ will contain a factor $[e^{-S(r)/2}]^N$, $S(r)$ being the
Euclidean action of the gravitational fluctuation, which is of
the order $(r/\ell_*)^2$. This is just an expression of the idea
that inside large fluctuations, interactions that involve a
large number of spacetime points are strongly suppressed. As the
size of the fluctuation decreases, the probability for events in
which three or more spacetime points are correlated increases,
in close analogy with the kinetic theory of gases: the higher
the density of molecules in the gas, the more probable is that a
large number of molecules collide at the same point. The
expansion parameter in this example is typically the density of
molecules. In our case, the natural expansion parameter is the
transition amplitude. It is given by the square root of the
two-point transition probability which in the semiclassical
approximation is of the form $e^{-S(r)}$.

Thus the $N$-local interaction term in ${\cal W}$ will be of
order $[e^{-S(r)/2}]^N$. In the weak-coupling approximation,
i.e., up to second order in the expansion parameter, the
trilocal and higher effective interactions do not contribute.
The terms corresponding to $N=0,1$ are local and can be absorbed
in the bare action (note that the coefficient $\upsilon$ is
constant and that the coefficients $\upsilon^{i_1}(t_1)$ cannot
depend on spacetime positions because of diffeomorphism
invariance). Consequently, we can write the action functional
${\cal W}$ as a bilocal whose most general form is \cite{fh65}
\begin{eqnarray}
{\cal W}[\phi,\varphi;t] \!\!\!&=&\!\!\! -\frac{1}{2}\int_0^t
ds\int_0^s ds'\{h_i[\phi;s]-h_i[\varphi;s]\}
\nonumber\\
\!\!\!&&\!\!\!\times \{\upsilon^{ij}(s-s')h_j[\phi;s']-
\upsilon^{ij}(s-s')^*h_j[\varphi;s']\}\,,
\end{eqnarray}
where we have renamed $\upsilon^{ij}(s,s')$ as
$\upsilon^{ij}(s-s')$, and without loss of generality we have
set $s>s'$. This complex coefficient is Hermitian in the pair of
indices $ij$ and depends on the spatial positions
$x_{\underline{i}}$ and $x_{\underline{j}}$ only through the
relative distance $|x_{\underline{i}}-x_{\underline{j}}|$. It is
of order $e^{-S(r)}$ and is concentrated within a spacetime
region of size $r$.

Let us now decompose $\upsilon^{ij}(\tau)$ in terms of its real
and imaginary parts as
\begin{equation}
\upsilon^{ij}(\tau) = c^{ij}(\tau)+i\dot f^{ij}(\tau)\,,
\end{equation}
where $c^{ij}(\tau)$ and $f^{ij}(\tau)$ are real and symmetric,
and the overdot denotes time derivative. The imaginary part is
antisymmetric in the exchange of $i,\tau$ and $j,-\tau$ and has
been written as a time derivative for convenience, since this
choice does not involve any restriction. The $f$ term can then
be integrated by parts to obtain
\begin{eqnarray}
{\cal W}[\phi,\varphi;t]\!\!\!&=&\!\!\! -\frac{1}{2}\int_0^t ds
\int_0^sds' c^{ij}(s-s')\{h_i[\phi;s]-h_i[\varphi;s]\}
\{h_j[\phi;s']- h_j[\varphi;s']\}
\nonumber\\
\!\!\!&&\!\!\!-\frac{i}{2}\int_0^t ds\int_0^s ds'
f^{ij}(s-s')\{h_i[\phi;s]-h_i[\varphi;s]\} \{\dot
h_j[\phi;s']+\dot h_j[\varphi;s']\}\,.
\end{eqnarray}
In this integration, we have ignored surface terms that
contribute, at most, to a finite renormalization of the bare
low-energy Hamiltonian.

The functions $f^{ij}(\tau)$ and $c^{ij}(\tau)$ characterize
spacetime foam in our effective description but, under fairly
general assumptions, the characterization can be carried out by
a smaller set of independent functions. In what follows we will
simplify this set. With this aim, we first write $f^{ij}(\tau)$
and $c^{ij}(\tau)$ in terms of their spectral counterparts
$\tilde f^{\underline{i}\underline{j}}(\omega)$ and $\tilde
c^{\underline{i}\underline{j}}(\omega)$. Lorentz invariance and
spatial homogeneity implies that $f^{ij}(\tau)$ and
$c^{ij}(\tau)$ must have the form
\begin{eqnarray}
f^{ij} (\tau)\!\!\!&=&\!\!\!\int_0^\infty d\omega \tilde
f^{\underline{i}\underline{j}}(\omega)
 8\pi \frac{\sin(\omega
|x_{\underline{i}}-x_{\underline{j}}|)} {\omega
|x_{\underline{i}}-x_{\underline{j}}|} \cos(\omega\tau)\,,
\\
c^{ij} (\tau)\!\!\!&=&\!\!\!\int_0^\infty d\omega \tilde
c^{\underline{i}\underline{j}}(\omega) 8\pi \frac{\sin(\omega
|x_{\underline{i}}-x_{\underline{j}}|)} {\omega
|x_{\underline{i}}-x_{\underline{j}}|} \cos(\omega\tau)\,,
\end{eqnarray}
for some real functions $\tilde
f^{\underline{i}\underline{j}}(\omega)$ and $\tilde
c^{\underline{i}\underline{j}}(\omega)$. It seems reasonable to
assume a kind of equanimity principle by which spacetime foam
produces interactions whose intensity does not depend on the
pair of interactions $h_i$ itself but on its independent
components for each mode, i.e., that the spectral interaction is
given by products of functions $\chi^{\underline{i}}(\omega)$:
\begin{eqnarray}
 \tilde f^{\underline{i}\underline{j}}(\omega)\!\!\!&=&\!\!\!
\chi^{\underline{i}}(\omega)\chi^{\underline{j}}(\omega)\,,
\\
\tilde c^{\underline{i}\underline{j}}(\omega)\!\!\!&=&\!\!\! g(\omega)
\chi^{\underline{i}}(\omega)\chi^{\underline{j}}(\omega)\,,
\end{eqnarray}
where $g(\omega)$ is a function that, together with
$\chi^{\underline{i}}(\omega)$, fully characterize spacetime
foam under these assumptions.

Then, $f^{ij}(\tau)$ and $c^{ij}(\tau)$ can be written as
\begin{eqnarray}
f^{ij} (\tau)\!\!\!&=&\!\!\!\int_0^\infty d\omega G^{ij}(\omega)
\cos(\omega\tau)\,,
\label{fij}\\
c^{ij}(\tau)\!\!\! &=&\!\!\!\int_0^\infty d\omega g(\omega)
G^{ij}(\omega)\cos(\omega\tau)\,,
\label{cij}
\end{eqnarray}
with
\begin{equation}
G^{ij}(\omega)=8\pi \frac{\sin(\omega
|x_{\underline{i}}-x_{\underline{j}}|)} {\omega
|x_{\underline{i}}-x_{\underline{j}}|}
\chi^{\underline{i}}(\omega)\chi^{\underline{j}}(\omega)\,.
\end{equation}
The functions $\chi^{\underline{i}}(\omega)$ can be interpreted
as the spectral effective couplings between spacetime foam and
low-energy fields. Since $\upsilon^{ij}(\tau)$ is of order
$e^{-S(r)}$ and is concentrated in a region of linear size $r$,
the couplings $\chi^{\underline{i}}(\omega)$ will have
dimensions of length, will be of order $e^{-S(r)/2}r$, and will
induce a significant interaction for all frequencies $\omega$ up
to the natural cutoff $r^{-1}$. On the other hand, the function
$g(\omega)$ has dimensions of inverse length and must be of
order $r^{-1}$. Actually, this function must be almost flat in
the frequency range $(0,r^{-1})$ to ensure that all the modes
contribute significantly to all bilocal interactions. As we will
see, the function $g(\omega)$ also admits a straightforward
interpretation in terms of the mean occupation number for the
mode of frequency $\omega$.

Once we have computed the influence functional $\mathcal{F}$, it
is possible to obtain the master equation that governs the
evolution of the density of low-energy fields, although we will
not follow this procedure here. We postpone the derivation of
the full master equation until next section.

The bilocal effective interaction does not lead to a unitary
evolution. The reason for this is that it is not sufficient to
know the fields and their time derivatives at an instant of time
in order to know their values at a later time: we need to know
the history of the system, at least for a time $r$.  There exist
different trajectories that arrive at a given configuration
$(\phi,\dot\phi)$. The future evolution depends on these past
trajectories and not only on the values of $\phi$ and $\dot
\phi$ at that instant of time. Therefore, the system cannot
possess a well-defined Hamiltonian vector field and suffers from
an intrinsic loss of predictability \cite{ew89}.

This can be easily seen if we restrict to the case in which
$f^{ij}(\tau)$ vanishes, i.e.,
$\upsilon^{ij}(\tau)=c^{ij}(\tau)$. Then, the influence
functional ${\cal F}_{\rm c}$ is the characteristic functional
of a Gaussian probability functional distribution, i.e., it can
be written as
\begin{equation}
{\cal F}_{\rm c}[\phi,\varphi;t]=\int {\cal D}\alpha
e^{-\frac{1}{2} \int_0^t ds \int_0^sds^\prime
\gamma_{ij}(s-s^\prime)\alpha^i(s)\alpha^j(s^\prime)}
 e^{i\int_0^t ds \alpha^i(s)\{ h_i[\phi;s]- h_i[\varphi;s]\}}\,.
\end{equation}
Here, the continuous matrix $\gamma_{ij}(s-s^\prime)$ is the
inverse of $c^{ij}(s-s^\prime)$, i.e.,
\begin{equation}
\int ds^{\prime\prime}\gamma_{ik}(s-s^{\prime\prime})
c^{kj}(s^{\prime\prime}-s^\prime)=\delta_i^j\delta(s-s^\prime)\,.
\end{equation}
Then, in this case, the propagator $\sso(t) $ has the form
\begin{equation}
\sso(t)=\int{\cal D}\alpha P[\alpha] \sso_\alpha(t)\,,
\end{equation}
where ${\sso}_\alpha(t)$ is just a factorizable propagator
associated with unitary evolution governed by the action
$S_0+\int\alpha^i h_i$ and
\begin{equation}
P[\alpha]=e^{-\frac{1}{2} \int_0^t ds\int_0^s ds^\prime
\gamma_{ij}(s-s^\prime)\alpha^i(s)\alpha^j(s^\prime)}\,.
\end{equation}
Therefore, $\sso(t) $ is just the average with Gaussian weight
$P[\alpha]$ of the unitary propagator $\sso_\alpha(t)$.

 Note that
the quadratic character of the distribution for the fields
$\alpha^i$ is a consequence of the weak-coupling approximation,
which keeps only the bilocal term in the action. Higher-order
terms would introduce deviations from this noise distribution.
The nonunitary nature of the bilocal interaction has been
encoded inside the fields $\alpha^i$, so that, when insisting on
writing the system in terms of unitary evolution operators, an
additional sum over the part of the system that is unknown
naturally appears. Note also that we have a different field
$\alpha^i$ for each kind of interaction $h_i$. Thus, we have
transferred the nonlocality of the low-energy field $\phi$ to
the set of fields $\alpha^i$, which are nontrivially coupled to
it and that represent spacetime foam.

\subsection{Semiclassical diffusion}
 We can see that the limit
of vanishing $f^{ij}(\tau)$, with nonzero $c^{ij}(\tau)$ (and
therefore real $\upsilon^{ij}(\tau)$), is a kind of
semiclassical approximation since, in this limit, one ignores
the quantum nature of the gravitational fluctuations. Indeed,
the fields $\alpha^i$ represent spacetime foam but, as we have
seen, the path integral for the whole system does not contain
any trace of the dynamical character of the fields $\alpha^i$.
It just contains a Gaussian probability distribution for them.
The path integral above can then be interpreted as a Gaussian
average over the classical noise sources $\alpha^i$.
Classicality here means that we can keep the sources $\alpha^i$
fixed, ignoring the noise commutation relations, and, at the end
of the calculations, we just average over them.

The low-energy density matrix $\rho$ then satisfies the
following master equation \cite{1ga98,2ga98,3ga98}
\begin{equation}
\dot \rho= -i \big[ H_0,\rho\big]- \int_0^\infty d\tau c^{ij}(\tau)
\big[h_{i},\big[h_{j}^{\mathsc I}(-\tau),\rho\big]\big]\,,
\end{equation}
where $h_j^{\mathsc I}(-\tau)=e^{-iH_0\tau}h_je^{iH_0\tau}$.
Since $e^{iH_0\tau}=1+O(\tau/l)$, the final form of the master
equation for a low-energy system subject to gravitational
fluctuations treated as a classical environment and at zeroth
order in $r/l$ (the effect of higher order terms in $r/l$ will
be thoroughly studied together with the quantum effects) is
\begin{equation}
\dot \rho= -i \big[ H_0,\rho\big]- \int_0^\infty  d\tau c^{ij}(\tau)
\big[h_{i},\big[h_{j},\rho\big]\big]
\end{equation}
(for similar approaches yielding this type of master equation
see also Refs. \cite{bs84,diosi87,percival95}).

The first term gives the low-energy Hamiltonian evolution that
would also be present in the absence of fluctuations. The second
term is a diffusion term which will be responsible for the loss
of coherence (and the subsequent increase of entropy). It is a
direct consequence of the foamlike structure of spacetime and
the related existence of a minimum length. Note there is no
dissipation term. This term is usually present in order to
preserve the commutation relations under time evolution.
However, we have considered the classical noise limit, i.e., the
noise $\alpha$ has been considered as a classical source and the
commutation relations are automatically preserved. We will see
that the dissipation term, apart from being of quantum origin,
is $r/l$ times smaller than the diffusion term and we have only
considered the zeroth order approximation in $r/l$.

The characteristic decoherence time $\tau_{d}$ induced by the
diffusion term can be easily calculated. Indeed, the interaction
Hamiltonian density $h_i$ is of order
$\ell_*^{-4}(\ell_*/l)^{2n_{\underline{i}}(1+s_{\underline{i}})}$
and $c^{ij}(\tau)$ is of order $e^{-S(r)}$. Furthermore, the
diffusion term contains one integral over time and two integrals
over spatial positions. The integral over time and the one over
relative spatial positions provide a factor $r^4$, since
$c^{ij}(\tau)$ is different from zero only in a spacetime region
of size $r^4$, and the remaining integral over global spatial
positions provides a factor $l^3$, the typical low-energy
spatial volume. Putting everything together, we see that the
diffusion term is of order
$l^{-1}\epsilon^2\sum_{{\underline{i}}{\underline{j}}}
(\ell_*/l)^{\eta_{\underline{i}}+\eta_{\underline{j}}}$, with
$\eta_{\underline{i}}=2n_{\underline{i}}(1+s_{\underline{i}})-2$
and $\epsilon=e^{-S(r)/2}(r/\ell_*)^2$. This quantity defines
the inverse of the decoherence time $\tau_d$. Therefore, the
ratio between the decoherence time $\tau_d$ and the low-energy
length scale $l$ is
\begin{equation}
\tau_d/l\sim
\epsilon^{-2}\bigg[\sum_{{\underline{i}}{\underline{j}}}
(\ell_*/l)^{\eta_{\underline{i}}+\eta_{\underline{j}}}
\bigg]^{-1}\,.
\end{equation}
Because of the exponential factor in $\epsilon$, only the
gravitational fluctuations whose size is very close to Planck
length will give a sufficiently small decoherence time. Slightly
larger fluctuations will have a very small effect on the
unitarity of the effective theory. For the interaction term that
corresponds to the mass of a scalar field, the parameter $\eta$
vanishes and, consequently, $\tau_d/l\sim
\epsilon^{-2}$. Thus, the scalar mass term will lose coherence
faster than any other interaction. Indeed, for higher spins
and/or powers of the field strength, $\eta\geq 1$ and therefore
$\tau_d/l$ increases by powers of $l/\ell_*$. For instance, the
next relevant decoherence time corresponds to the scalar-fermion
interaction term $\phi^2\bar\psi\psi$, which has an associated
decoherence ratio $\tau_d/l\sim \epsilon^{-2}l/\ell_*$. We see
that the decoherence time for the mass of scalars is independent
of the low-energy length scale and, for gravitational
fluctuations of size close to Planck length, $\epsilon$ may be
not too small so that scalar masses may lose coherence fairly
fast, maybe in a few times the typical evolution scale. Hawking
has argued \cite{ha96} that this might be the reason for not
observing the Higgs particle. Higher power and/or spin
interactions will lose coherence much slower but for
sufficiently high energies $l^{-1}$, although much smaller than
the gravitational fluctuations energy $r^{-1}$, the decoherence
time may be small enough. This means that quantum fields will
lose coherence faster for higher-energy regimes. Hawking has
also suggested that loss of quantum coherence might be
responsible for the vanishing of the $\theta$ angle in quantum
chromodynamics \cite{ha96}.

\subsection{Spacetime foam as a quantum bath}
 As we have briefly mentioned before, considering that the
coefficients $\upsilon^{ij}$ are real amounts to ignore the
quantum dynamical nature of spacetime foam, paying attention
only to its statistical properties. In what follows, we will
study these quantum effects and show that spacetime foam can be
effectively described in terms of a quantum thermal bath with a
nearly Planckian temperature that has a weak interaction with
low-energy fields. As a consequence, other effects, apart from
loss of coherence, such as Lamb and Stark transition-frequency
shifts, and quantum damping, characteristic of systems in a
quantum environment \cite{ga91,ca93}, naturally appear as
low-energy predictions of this model \cite{1ga98,2ga98,3ga98}.

 Let us consider a Hamiltonian of the
form
\begin{equation}
H=H_0+H_{\rm int}+H_{\rm b}\,.
\end{equation}
$H_0$ is the bare Hamiltonian that represents the low-energy
fields and $H_{\rm b}$ is the Hamiltonian of a bath that, for
simplicity, will be represented by a real massless scalar field.
The interaction Hamiltonian will be of the form $H_{\rm
int}=\xi^i h_i$, where the noise operators $\xi^i$ are given by
\begin{equation}
\xi^{\underline{i}}(x,t)= \int dx'
\chi^{{\underline{i}}}(x-x')p(x',t)\,.
\end{equation}
Here, $p(x,t)$ is the momentum of the bath scalar field whose
mode decomposition has the form
\begin{equation}
p(x,t)=i\int dk \sqrt \omega [ a^\dag(k) e^{i(\omega t-k
x)}-a(k) e^{-i(\omega t-k x)}]\,,
\end{equation}
$\omega=\sqrt{k^2}$, and $a$ and $a^\dag$ are, respectively, the
annihilation and creation operators associated with the bath;
$\chi^{\underline{i}}(y)$ represent the couplings between the
low-energy field and the bath in the position representation.
Since we are trying to construct a model for spacetime foam, we
will assume that the couplings $\chi^{{\underline{i}}}(y)$ will
be concentrated on a region of radius $r$ and that they are
determined by the spectral couplings
$\chi^{{\underline{i}}}(\omega)$ introduced before:
\begin{equation}
\chi^{\underline{i}}(y)=\int \frac{dk}{\omega}
\chi^{{\underline{i}}}(\omega) \cos (ky)\,.
\end{equation}

The influence functional in this case has the form \cite{fh65}
\begin{equation}
{\cal F}[\phi,\varphi;t]=\int Dq' DQ' \rho_{\rm b}[q',Q';0]
\int{\cal D} q {\cal D} Q e^{i\{S_{\rm b}[q;t]-S_{\rm b}[Q;t]\}}
e^{i\{S_{\rm int}[\phi,q;t]-S_{\rm int}[\varphi,Q;t]\}}\,,
\end{equation}
where these path integrals are performed over paths $q(s)$ and
$Q(s)$ such that at the initial time match the values $q'$ and
$Q'$ and $S_{\rm b}$ is the action of the bath.

If we assume that the bath is in a stationary, homogeneous, and
isotropic state, this influence functional can be computed to
yield an influence action ${\cal W}$ of the form discussed
above. Furthermore, for a thermal state with temperature $T\sim
1/r$, the function $g(\omega)$ has the form
\begin{equation}
g(\omega)=\omega[N(\omega)+1/2]\,,
\end{equation}
where $N(\omega)=[\exp(\omega/T)-1]^{-1}$ is the mean occupation
number of the quantum thermal bath corresponding to the
frequency $\omega$. Recall that the functions $G^{ij}(\omega)$
and, hence, $f^{ij}(\tau)$ are uniquely determined by the
couplings $\chi^{\underline{i}}(\omega)$. In particular, they
are completely independent of the state of the bath or the
system. All the relevant information about the bath is encoded
in the function $g(\omega)$.

With this procedure, we see that spacetime foam can be
represented by a quantum bath determined by $g(\omega)$ that
interacts with the low-energy fields by means of the couplings
$\chi^{\underline{i}}(\omega)$ which characterize spacetime
foam, in the sense that both systems produce the same low-energy
effects.

This model that we have proposed is particularly suited to the
study of low-energy effects produced by simply connected
topology fluctuations such as closed loops of virtual black
holes \cite{ha96}. Virtual black holes will not obey classical
equations of motion but will appear as quantum fluctuations of
spacetime and thus will become part of the spacetime foam as we
have discussed. Particles could fall into these black holes and
be re-emitted. The scattering amplitudes of these processes
\cite{ha96,hr97} could be interpreted as being produced by
nonlocal effective interactions that would take place inside the
fluctuations and the influence functional obtained above could
then be interpreted as providing the evolution of the low-energy
density matrix in the presence of a bath of ubiquitous quantum
topological fluctuations of the virtual-black-hole type.

\subsection{Wormholes and coherence}
Euclidean solutions of the wormhole type were obtained for a
variety of matter contents (see, e.g.,
\cite{giddings88a,halliwell89,keay89}). Quantum solutions to the
Wheeler-DeWitt equation that represent wormholes can be found in
Refs. \cite{ha88,hp90,lyons89,dowker90,dowker91,barcelo96}.
These solutions allowed the calculation of the effective
interactions that they introduce in low-energy physics
\cite{ha88,lyons89,dowker90,dowker91,barcelo98a,barcelo98b,lukas95}.

Wormholes do not seem to induce loss of coherence despite the
fact that they render spacetime multiply connected
\cite{co88,giddings88b}. The reason why they seem to preserve
coherence is that, in the dilute gas approximation, they join
spacetime regions that may be far apart from each other and
therefore, both wormhole mouths must be delocalized, i.e., the
multiply-connectedness requires energy and momentum conservation
in both spacetime regions separately. In this way, wormholes can
be described as bilocal interactions whose coefficients
$\upsilon^{ij}$ do not depend on spacetime positions.
Diffeomorphism invariance on each spacetime region also requires
the spacetime independence of $\upsilon^{ij}$. This can also be
seen by analyzing these wormholes from the point of view of the
universal covering manifold, which is, by definition, simply
connected. Here, each wormhole is represented by two boundaries
located at infinity and suitably identified. This identification
is equivalent to introducing coefficients $\upsilon^{ij}$ that
relate the bases of the Hilbert space of wormholes in both
regions of the universal covering manifold. Since
$\upsilon^{ij}$ are just the coefficients in a change of basis,
they will be constant. As a direct consequence, the correlation
time for the fields $\alpha^i$ is infinite. This means that the
fields $\alpha^i$ cannot be interpreted as noise sources that
are Gaussian distributed at each spacetime point independently.
Rather, they are infinitely coherent thus giving rise to
superselection sectors. The Gaussian distribution to which they
are subject is therefore global, spacetime independent. The only
effect of wormholes is thus introducing local interactions with
unknown but spacetime independent coupling constants
\cite{co88,giddings88b,hawking88}. The spacetime independence
implies that, once a experiment to determine one such constant
is performed, it will remain with the obtained value forever, in
sharp contrast with those induced by simply-connected
topological fluctuations such as virtual black holes. In this
way and because of the infinite-range correlations induced by
wormholes, which forbid the existence of asymptotic regions
necessary to analyze scattering processes, the loss of coherence
produced by these fluctuations should actually be ascribed to
the lack of knowledge of the initial state or, in other words,
to the impossibility of preparing arbitrarily pure quantum
states \cite{co88}.

One could also expect some effects originated in their quantum
nature. However, the coefficients $\upsilon^{ij}$ are spacetime
independent. This means that $c^{ij}$ are constant and,
consequently, $\tilde c^{\underline{i}\underline{j}}(\omega)\sim
\delta(\omega)$. As we have argued, $\tilde
c^{\underline{i}\underline{j}}(\omega)$ and $\tilde
f^{\underline{i}\underline{j}}(\omega)$ are related by a nearly
flat function so that $\tilde
f^{\underline{i}\underline{j}}(\omega)\sim\delta(\omega)$ as
well. This in turn implies that $f^{ij}$ is also constant and
$\dot f^{ij}=0$, therefore concluding that $\upsilon^{ij}$ is
real. We have already argued that in the case of real
$\upsilon^{ij}$, no quantum effects will show up.

Wormhole spacetimes do not lead, strictly speaking, to loss of
quantum coherence although global hyperbolicity does not hold.
On the other hand, the difficulties in quantum gravity with
unitary propagation mainly come from the quantum field theory
axiom of asymptotic completeness \cite{ha82,alvarez83}, which is
closely related to global hyperbolicity. Indeed, in order to
guarantee asymptotic completeness, it is necessary that the
expectation value of the fields at any spacetime position be
determined by their values at a Cauchy surface at infinity.
Topologically nontrivial spacetimes however are not globally
hyperbolic in general and therefore do not admit a foliation in
Cauchy surfaces. Let us have a closer look at this issue.

Gravitational entropy, which is closely related to the loss of
quantum coherence, has its origin in the existence of
two-dimensional spheres in Euclidean space that cannot be
homotopically contracted to a point, i.e., with nonvanishing
second Betti number. These two-dimensional surfaces become fixed
points of the timelike Killing vector, so that global
hyperbolicity is lost. A well-known example (for other more
sophisticated examples see Ref. \cite{hawking98}) is a
Schwarzschild black hole whose Euclidean sector is described by
the metric
\begin{equation}
ds^2=f(r)dt^2+f(r)^{-1} dr^2+r^2 d\Omega_2^2\,,
\end{equation}
with $f(r)=1-2\ell_*^2m/r$, $m$ being the black hole mass. In
order to make this solution regular, we consider the region
$r\geq 2\ell_*^2m$ and set $t$ to be periodic with period
$\beta=8\pi \ell_*^2m$. The surface defined by $r=2\ell_*^2m$ is
a fixed point of the Killing vector $\partial_t$. Thus, we have
a spacetime with the topology of $\Re^2\times S^2$, so that
$B_2=1$. As we will see below, it is the existence of this
surface that accounts for the entropy of this spacetime. This
does not mean that it is localized in the surface itself.
Rather, it is a global quantity characteristic of the whole
spacetime manifold. The Euclidean action of this solution is
given by the sum of the contributions $I_{\rm fp}$, $I_{\infty}$
of the two surface terms at $r=2\ell_*^2m$ and $r=\infty$. In
the semiclassical approximation, the partition function is given
by $Z=e^{-I_{\rm fp}-I_{\infty}}$. Taking into account that the
entropy is $S=\ln Z-\beta E$ and that $\beta E$ is precisely the
surface term at infinity $\beta E=-I_{\infty}$, we conclude that
the entropy is given by the surface term at $r=2\ell_*^2m$,
$S=-I_{\rm fp}=4\pi\ell_*^2 m^2$, as is well-known.

In the wormhole case, the second Betti number is zero and the
first and third Betti numbers are equal. For a spacetime with a
single wormhole, $B_1=B_3=1$. This means that there exists one
circle that cannot be homotopically contracted to a point and
that there also exists one three-sphere that is not homotopic to
a point, but all two-spheres are contractible. Regular solutions
of this sort can be identified with the wormhole throat. The
only contributions to the Euclidean action in this case come
from the asymptotic regions, which is precisely the term that we
have to subtract from $\ln Z$, in the semiclassical
approximation, in order to calculate the gravitational entropy.
Thus, wormholes have vanishing entropy despite the fact that
they are not globally hyperbolic. From the point of view of
their universal covering manifold, a wormhole is represented by
two three-surfaces whose contribution to the action are equal in
absolute value but with opposite sign because of their reverse
orientations, thus leaving only the asymptotic contribution,
irrelevant as far as the entropy is concerned (for a different
approach see Ref. \cite{gonzalez91}). The striking difference
between wormholes and virtual black holes is precisely the
formation of horizons which has no counterpart in the wormhole
case. This is closely related to the issue of the infinite-range
spacetime correlations established by wormholes versus the
finite size of the regions occupied by virtual black holes or
quantum time machines, for instance.


\section{Low-energy effective evolution}
\indent

As we have already mentioned, from the influence functional
obtained in the previous section, we can obtain the master
equation satisfied by the low-energy density matrix, although
here we will follow a different procedure: We will derive the
master equation in the canonical formalism from von Neumann
equation for the joint system of the low-energy fields plus the
effective quantum bath coupled to them that accounts for the
effects of spacetime foam.

\subsection{Master equation}
It is easy to see that the function $f^{ij}(\tau)$ given in Eq.
(\ref{fij}) determines the commutation relations at different
times of the noise variables. Indeed, taking into account the
commutation relations for the annihilation and creation
operators $a$ and $a^\dag$, i.e.,
\begin{equation}
\big[a(k),a(k')\big]=0\,,
\hspace{5mm}
\big[a(k),a^\dag(k')\big]=\delta(k-k')\,,
\end{equation}
we obtain by direct calculation the relation
\begin{equation}
\big[\xi^i(t),\xi^j(t')\big]=i \frac{d}{dt} f^{ij}(t-t')\,,
\end{equation}

Similarly, the function $c^{ij}(\tau)$ of Eq. (\ref{cij})
determines the average of the anticommutator of the noise
variables,
\begin{equation}
\frac{1}{2}\big\langle\big[\xi^i(t),\xi^j(t')\big]_+
\big\rangle=c^{ij}(t-t')\,,
\end{equation}
where the average of any operator $Q$ has been defined as
$\langle Q\rangle\equiv {\rm tr}_{\rm b}(Q\rho_{\rm b})$,
provided that the bath is in a stationary, homogeneous, and
isotropic state determined by $g(\omega)$, i.e.,
\begin{equation}
\langle a(k)\rangle=0\,,
\hspace{5mm}
\langle a(k)a(k')\rangle=0\,,
\hspace{5mm}
\langle a^\dag(k)a(k')\rangle=[g(\omega)/\omega-1/2]\delta(k-k')\,.
\end{equation}

We are now ready to write down the master equation for the
low-energy density matrix. We will describe the whole system
(low-energy field and bath) by a density matrix $\rho_{\mathsc
T}(t)$. We will assume that, initially, the low energy fields
and the bath are independent, i.e., that at the time $t=0$
\begin{equation}
\rho_{\mathsc T}(0)=\rho(0)\otimes \rho_{\rm b}\,.
\end{equation}
If the low-energy fields and the bath do not decouple at any
time, an extra renormalization term should be added to the
Hamiltonian. In the interaction picture, the density matrix has
the form
\begin{equation}
\rho_{\mathsc T}^{\mathsc I}(t)=U^\dag(t)\rho_{\mathsc T}(t)U(t)\,,
\end{equation}
with $U(t)=U_0(t)U_{\rm b}(t)$, where $U_0(t)= e^{-iH_0t}$ and
$U_{\rm b}(t)=e^{-iH_{\rm b}t}$. It obeys the equation of motion
\begin{equation}
\dot\rho_{\mathsc T}^{\mathsc I}(t)=-i \big[\xi^i(t) h_i^{\mathsc
I}(t),\rho_{\mathsc T}^{\mathsc I}(t)\big]\,.
\end{equation}
Here,
\begin{eqnarray}
\xi^i(t)\!\!\!&=&\!\!\!U^\dag(t)\xi^iU(t)=
U_{\rm b}^\dag(t)\xi^iU_{\rm b}(t)\,,
\\
h_i^{\mathsc
I}(t)\!\!\!&=&\!\!\!U^\dag(t)h_iU(t)=U_0^\dag(t)h_iU_0(t)\,.
\end{eqnarray}
Integrating this evolution equation and introducing the result
back into it, tracing over the variables of the bath, defining
$\rho^{\mathsc I}(t)\equiv{\rm tr}_{\rm b}[\rho_{\mathsc
T}^{\mathsc I}(t)]$, and noting that ${\rm tr}_{\rm
b}[\xi^i(t)h_i^{\mathsc I}(t)\rho_{\mathsc T}^{\mathsc
I}(t_0)]=0$, we obtain
\begin{equation}
\dot\rho^{\mathsc I}(t)= -\int_{t_0}^t  dt' {\rm tr}_{\rm
b}\left\{\big[ \xi^i(t) h_i^{\mathsc I}(t), \big[\xi^j(t')
h_j^{\mathsc I}(t'),\rho_{\mathsc T}^{\mathsc
I}(t')\big]\big]\right\}\,.
\end{equation}

In the weak-coupling approximation, which implies that
$\xi^ih_i$ is much smaller than $H_0$ and $H_{\rm b}$ (this is
justified since it is of order $\epsilon$), we assume that the
bath density matrix does not change because of the interaction,
so that $\rho_{\mathsc T}^{\mathsc I}(t)=\rho^{\mathsc
I}(t)\otimes\rho_{\rm b}$. The error introduced by this
substitution is of order $\epsilon$ and ignoring it in the
master equation amounts to keep terms only up to second order in
this parameter. Since $\big[\xi^i(t), h_j^{\mathsc
I}(t')\big]=0$ because $\big[\xi^i, h_j\big]=0$, the right hand
side of this equation can be written in the following way
\begin{equation}
-\int_{0}^t dt'\big\{ c^{ij}(t-t') \big[h_i^{\mathsc
I}(t),\big[h_j^{\mathsc I}(t'),\rho^{\mathsc I}(t')\big]\big]
+\frac{i}{2}f^{ij}(t-t') \big[h_i^{\mathsc
I}(t),\big[h_j^{\mathsc I}(t'),\rho^{\mathsc
I}(t')\big]_+\big]\big\}\,.
\end{equation}

The Markov approximation allows the substitution of
$\rho^{\mathsc I}(t')$ by $\rho^{\mathsc I}(t)$ in the master
equation because the integral over $t'$ will get a significant
contribution from times $t'$ that are close to $t$ due to the
factors $\dot f^{ij}(t-t')$ and $c^{ij}(t-t')$ and because, in
this interval of time, the density matrix $\rho^{\mathsc I}$
will not change significantly. Indeed, the typical evolution
time of $\rho^{\mathsc I}$ is the low-energy time scale $l$,
which will be much larger than the time scale $r$ associated
with the bath. If we perform a change of the integration
variable from $t'$ to $\tau=t-t'$, write
\begin{equation}
\rho^{\mathsc I}(t')=\rho^{\mathsc I}(t-\tau)=
\rho^{\mathsc I}(t)-\tau\dot\rho^{\mathsc I}(t) +O(\tau^2)\,,
\end{equation}
and introduce this expression in the master equation above, we
easily see that the error introduced by the Markovian
approximation is of order $\epsilon^2$, i.e., it amounts ignore
a term of order $\epsilon^4$. The upper integration limit $t$ in
both integrals can be substituted by $\infty$ for evolution
times $t$ much larger than the correlation time $r$, because of
the factors $\dot f^{ij}(\tau)$ and $c^{ij}(\tau)$ that vanish
for $\tau>r$.

Then, after an integration by parts of the $f$ term, and
transforming the resulting master equation back to the
Schr\"{o}dinger picture we obtain
\begin{equation}
\dot\rho= -i\big[H_0',\rho\big]-\frac{i}{2}\int_0^\infty d\tau
f^{ij}(\tau) \big[h_i,\big[\dot h_j^{\mathsc
I}(-\tau),\rho\big]_+\big] -\int_0^\infty d\tau c^{ij}(\tau)
\big[h_i,\big[h_j^{\mathsc I}(-\tau),\rho\big]\big]\,,
\end{equation}
where $ H_0'=H_0-\frac{1}{2}f^{ij}(0) h_ih_j$ is just the
original low-energy Hamiltonian plus a finite renormalization
originated in the integration by parts of the $f$ term. It can
be checked that the low-energy density matrix $\rho(t)$ obtained
by means of the influence functional $\mathcal{F}$ is indeed a
solution of this master equation.

Before discussing this equation in full detail, let us first
study the classical noise limit. With this aim, let us introduce
the parameter
\begin{equation}
\sigma=\int dk' \big[a(k),a^\dag(k')\big]\,,
\end{equation}
which is equal to 1 for quantum noise and 0 for classical noise.
Then, the $f$ term is proportional to $\sigma$ and therefore
vanishes in the classical noise limit. On the other hand, the
function $g(\omega)$ becomes $g(\sigma \omega)$ when introducing
the parameter $\sigma$. In the limit $\sigma\rightarrow 0$, it
acquires the value $g(0)$ which is a constant of order $1/r$.
Therefore, $c^{ij}(\tau)$ becomes in this limit $c_{\rm
class}^{ij}(\tau)=g(0)f^{ij}(\tau)$. Also, the renormalization
term of the low-energy Hamiltonian vanishes in this limit. In
this way, we have arrived at the same master equation that we
obtained in the previous section. This is not surprising because
the origin of the $f$ term is precisely the noncommutativity of
the noise operators, i.e., its quantum nature, while the $c_{\rm
class}$ term actually contains the information about the state
of the bath. In the case of a thermal bath, $g(0)$ is precisely
the temperature of the bath. At zeroth order in $r/l$, the
master equation for classical noise then acquires the form
\begin{equation}
\dot \rho= -i \big[ H_0,\rho\big]- \int_0^\infty  d\tau c_{\rm
class}^{ij}(\tau) \big[h_{i},\big[h_{j},\rho\big]\big]\,.
\end{equation}

\subsection{Low-energy effects}
Let us now analyze the general master equation, valid up to
second order in $\epsilon$, that takes into account the quantum
nature of the gravitational fluctuations. These contributions
will be fairly small in the low-energy regime, but may provide
interesting information about the higher-energy regimes in which
$l$ may be of the order of a few Planck lengths and for which
the weak-coupling approximation is still valid. In order to see
these contributions explicitly, let us further elaborate the
master equation. In terms of the operator $L_0$ defined as
$L_0\cdot A=\big[H_0,A\big]$ acting of any low-energy operator
$A$, the time dependent interaction $h^{\mathsc I}_j(-\tau)$ can
be written as
\begin{equation}
h^{\mathsc I}_j(-\tau)=e^{-iL_0\tau}h_j\,.
\end{equation}
The interaction $h_j$ can be expanded in eigenoperators
$h_{j\Omega}^{\pm}$ of the operator $L_0$, i.e.,
\begin{equation}
h_j=\int d\mu_\Omega\left(h_{j\Omega}^++h_{j\Omega}^-\right)\,,
\end{equation}
with $L_0\cdot h_{j\Omega}^{\pm}=\pm \Omega h_{j\Omega}^{\pm}$ and
$d\mu_\Omega$ being an appropriate spectral measure, which is
naturally cut off around the low-energy scale $l^{-1}$. This
expansion always exists provided that the eigenstates of $H_0$
form a complete set. Then, $h^{\mathsc I}_j(-\tau)$ can be
written as
\begin{equation}
h^{\mathsc I}_j(-\tau)=\int d\mu_\Omega (e^{-i\Omega \tau }
h^+_{j\Omega }+e^{i\Omega \tau } h^-_{j\Omega })\,.
\end{equation}
It is also convenient to define the new interaction operators
for each low-energy frequency $\Omega$
\begin{equation}
h^1_{j\Omega}=h^+_{j\Omega }-h^-_{j\Omega }\,,
\hspace{5mm}
h^2_{j\Omega}=h^+_{j\Omega }+h^-_{j\Omega }\,.
\end{equation}

The quantum noise effects are reflected in the master equation
through the term proportional to $f^{ij}(\tau)$ and the term
proportional to $c^{ij}(\tau)$, both of them integrated over
$\tau\in (0,\infty)$. Because of these incomplete integrals,
each term provides two different kinds of contributions whose
origin can be traced back to the well-know formula
\begin{equation}
\int_0^\infty d\tau e^{i\omega\tau} =\pi\delta(\omega)+{\cal
P}(i/\omega)\,,
\end{equation}
where ${\cal P}$ is the Cauchy principal part \cite{rs72}.

The master equation can then be written in the following form
\begin{equation}
\dot \rho =-(iL_0'+L_{\rm diss}+L_{\rm diff}+
iL_{\rm s-l})\cdot\rho\,,
\end{equation}
where the meaning of the different terms are explained in what
follows.

The first term $-iL_0'\cdot\rho$, with
$L_0'\cdot\rho=\big[H_0',\rho\big]$, is responsible for the
renormalized low-energy Hamiltonian evolution. The
renormalization term is of order $\varepsilon^2$ as compared
with the low-energy Hamiltonian $H_0$, where $\varepsilon^2=
\epsilon^2\sum_{{\underline{i}}{\underline{j}}}
(\ell_*/l)^{\eta_{\underline{i}}+\eta_{\underline{j}}}$ and,
remember, $\eta_{\underline{i}}= 2n_{\underline{i}}
(1+s_{\underline{i}})-2$ is a parameter specific to each kind of
interaction term $h_i$.

The dissipation term
\begin{equation}
L_{\rm diss}\cdot\rho=\frac{\pi}{4}\int d\mu_\Omega \Omega
G^{ij}(\Omega) \big[ h_i,\big[h^1_{j\Omega},\rho\big]_+\big]
\end{equation}
is necessary for the preservation in time of the low-energy
commutators in the presence of quantum noise. As we have seen,
it is proportional to the commutator between the noise creation
and annihilation operators and, therefore, vanishes in the
classical noise limit. Its size is of order
$\varepsilon^2r/l^2$.

The diffusion process is governed by
\begin{equation}
L_{\rm diff}\cdot\rho=\frac{\pi}{2}\int d\mu_\Omega g(\Omega)
G^{ij}(\Omega)
\big[h_i,\big[h^2_{j\Omega},\rho\big]\big]\,,
\end{equation}
which is of order $\varepsilon^2/l$.

The next term provides an energy shift which can be interpreted
as a mixture of a gravitational ac Stark effect and a Lamb shift
by comparison with its quantum optics analog \cite{ga91,ca93}.
Its expression is
\begin{equation}
L_{\rm s-l} =-\int d\mu_\Omega {\cal P}
\int_{0}^{\infty } d\omega\frac{\Omega }{\omega^2-
\Omega^2 } G^{ij}(\omega)
\left\{g(\omega) \big[h_i,\big[h^1_{j\Omega},\rho\big]\big]
+\frac{\Omega}{2}
\big[h_i,\big[h^2_{j\Omega},\rho\big]_+\big]\right\}\,.
\end{equation}
The second term is of order $\varepsilon^2r^2/l^3$, which is
fairly small. However, the first term will provide a significant
contribution of order $\varepsilon^2r/l^2[\ln(l/r)+1]$. This
logarithmic dependence on the relative scale is indeed
characteristic of the Lamb shift \cite{ga91,ca93,it85}. As we
have argued the function $g(\omega)$ must be fairly flat in the
whole range of frequencies up to the cutoff $1/r$ and be of
order $1/r$ in order to reproduce the appropriate correlations
$c^{ij}(\tau)$. A thermal bath, for instance, produces a
function $g(\omega)$ with the desired characteristics, at least
at the level of approximation that we are considering. In this
specific case, it can be seen that the logarithmic contribution
to the energy shift is not present and it would only appear in
the zero temperature limit. However, since we are modeling
spacetime foam with this thermal bath, the effective temperature
is $1/r$, which is close to Planck scale and certainly far from
zero. From the practical point of view, the presence or not of
this logarithmic contribution is at most an order of magnitude
larger than the standard one and therefore it does not
significantly affect the results. Almost any other state of the
bath with a more or less uniform frequency distribution will
contain such logarithmic contribution.

As a summary, the $f$ term provides a dissipation part,
necessary for the preservation of commutators, and a fairly
small contribution to what can be interpreted as a gravitational
Lamb shift. On the other hand, the $c$ term gives rise to a
diffusion term and a shift in the oscillation frequencies of the
low-energy fields that can be interpreted as a mixture of a
gravitational Stark effect and a Lamb shift. The size of these
effects, compared with the bare evolution, are the following:
the diffusion term is of order $\varepsilon^2$ (see, however,
Refs. \cite{ellis97,ellis97b}); the damping term is smaller by a
factor $r/l$, and the combined effect of the Stark and Lamb
shifts is of order $(r/l)[\ln(l/r)+1]$ as compared with the
diffusion term. Note that the quantum effects induced by
spacetime foam become relevant as the low-energy length scale
$l$ decreases, as we see from the fact that these effects depend
on the ratio $r/l$, while, in this situation, the diffusion
process becomes faster, except for the mass of scalars, which
always decoheres in a time scale which is close to the
low-energy evolution time.

\subsection{Observational and experimental prospects}
These quantum gravitational effects are just energy shifts and
decoherence effects similar to those appearing in other areas of
physics, where fairly well established experimental procedures
and results exist, and which can indeed be applied here,
provided that sufficiently high accuracy can be achieved.

Neutral kaon beams have been proposed as experimental systems
for measuring loss of coherence owing to quantum gravitational
fluctuations \cite{el84,huet95,huet96,bf98}. In these systems,
the main experimental consequence of the diffusion term
(together with the dissipative one necessary for reaching a
stationary regime) is violation of CPT \cite{ha82,pa82} because
of the nonlocal origin of the effective interactions (see also
Refs. \cite{fivel97,ahluwalia99}). The estimates for this
violation are very close to the values accessible by current
experiments with neutral kaons and will be within the range of
near-future experiments. Macroscopic neutron interferometry
\cite{el84,ze84} provides another kind of experimental systems
in which the effects of the diffusion term may have measurable
consequences since they may cause the disappearance of the
interference fringes \cite{el84,ze84}.

As for the gravitational Lamb and Stark effects, they are energy
shifts that depend on the frequency, so that different
low-energy modes will undergo different shifts. This translates
into a modification of the dispersion relations, which makes the
velocity of propagation frequency-dependent, as if low-energy
fields propagated in a ``medium''. Therefore, upon arrival at
the detector, low-energy modes will experience different time
delays (depending on their frequency) as compared to what could
be expected in the absence of quantum gravitational
fluctuations. These time delays in the detected signals will be
very small in general. However, it might still be possible to
measure them if we make the low-energy particles travel large
(cosmological) distances. In fact, $\gamma$-ray bursts provide
such a situation as has been recently pointed out \cite{am98}
(see also Refs. \cite{am98b,schaefer98,biller98}), thus opening
a new doorway to possible observations of these quantum
gravitational effects. These authors assume that the dispersion
relation for photons has a linear dependence on $r/l$ because of
quantum gravitational fluctuations, i.e., that the speed of
light is of the form $v\sim 1+\zeta r/l$, with $\zeta$ being an
unknown parameter of order 1 (see also Ref. \cite{gambini98}).
In this situation, photons that travel a distance $L$ will show
a frequency-dependent time delay $\Delta t\sim \zeta L r/l$.
Using data from a $\gamma$-ray flare associated with the active
galaxy Markarian 421 \cite{biller98,gaidos96} which give
$l^{-1}\sim 1$TeV, $L\sim 1.1\times 10^{16}$ light-seconds, and
a variability time scale $\delta t$ less than 280 seconds, it
can be obtained the upper bound $\zeta r/\ell_*<250$. If $\zeta$
is indeed of order 1, this inequality implies an upper limit on
the scale $r$ of the gravitational fluctuations of a few hundred
Planck lengths. One would then expect that the presence of the
gravitational Lamb and Stark shifts predicted above could be
observationally tested. However, in spacetime foam the role of
the parameter $\zeta$ is played by $\varepsilon^2$ and this
quantity is much smaller than 1, since it contains two factors
which are smaller than 1 for different reasons. The first one is
$e^{-S(r)}(r/\ell_*)^2$. In the semiclassical approximation to
nonperturbative quantum gravity, this exponential can be
interpreted as the density of topological fluctuations of size
$r$, which decreases with $r$ fairly fast. The second factor is,
for the electromagnetic field, of the form $(\ell_*/l)^4$; it
comes from the spin dependence of the effective interactions and
is closely related to the existence of a length scale in quantum
gravity. Then, $\varepsilon^2$ in this case maybe so small that
might render any bound on the size of quantum spacetime foam
effects on the electromagnetic field nonrestrictive at all.


\section{Real clocks}
\indent

In previous sections, we have analyzed the evolution of
low-energy fields in the bath of quantum gravitational
fluctuations that constitute spacetime foam. Here we will
briefly discuss the evolution of physical systems when measured
by real clocks, which are generally subject to errors and
fluctuations, in contrast with ideal clocks which, although
would accurately measure the time parameter that appears in the
Schr\"{o}dinger equation, do not exist in nature (see, e.g., Refs.
\cite{wigner57,salecker58,peres80,page83,hartle88,unruh89b}).
The evolution according to real clocks bears a close resemblance
with low-energy fields propagating in spacetime foam, although
there also exist important differences which will be discussed
at the end of this section.

Quantum real clocks inevitably introduce uncertainties in the
equations of motion, as has been widely discussed in the
literature from various points of view (see, e.g., Refs.
\cite{wigner57,salecker58,peres80,page83,hartle88,unruh89b}).
Actually, real clocks are not only subject to quantum
fluctuations. They are also subject to classical imperfections,
small errors, that can only be dealt with statistically. For
instance, an unavoidable classical source of stochasticity is
temperature, which will introduce thermal fluctuations in the
behavior of real clocks. Thus, the existence of ideal clocks is
also forbidden by the third law of thermodynamics. Even at
zero-temperature, the quantum vacuum fluctuations of quantum
field theory make propagating physical systems (real clocks
among them) suffer a cold-diffusion and consequently a need for
a stochastic description of their evolution \cite{gour98}.

Let us study, within the context of the standard quantum theory,
the evolution of an arbitrary system according to a real clock
\cite{egusquiza98}.

\subsection{Good real clocks}
A real clock will be a system with a degree of freedom $t$ that
closely follows the ideal time parameter $t_{\rm i}$, i.e.,
$t_{\rm i}=t+\Delta(t)$, where $\Delta(t)$ is the error at the
real-clock time $t$. Given any real clock, its characteristics
will be encoded in the probability functional distribution for
the continuous stochastic processes $\Delta(t)$
\cite{kampen81,gardiner85} of clock errors, ${\cal
P}[\Delta(t)]$, which must satisfy appropriate conditions, so
that it can be regarded as a good clock.

A first property is that Galilean causality should be preserved,
i.e., that causally related events should always be properly
ordered in clock time as well, which implies that $t_{\rm i}
(t')>t_{\rm i}(t)$ for every $t'>t$. In terms of the derivative
$\alpha(t)=d\Delta(t)/dt$ of the stochastic process $\Delta(t)$,
we can state this condition as requiring that, for any
realization of the stochastic sequence, $\alpha(t)>-1$.

A second condition that we would require good clocks to fulfill
is that the expectation value of relative errors, determined by
the stochastic process $\alpha(t)$, be zero, i.e.,
$\langle\alpha(t)\rangle=0$ for all $t$. Furthermore, a good
clock should always behave in the same way (in a statistical
sense). We can say that the clock behaves consistently in time
as a good one if those relative errors $\alpha(t)$ are
statistically stationary, i.e., the probability functional
distribution ${\cal P}[\alpha(t)]$ for the process of relative
errors $\alpha(t)$ (which can be obtained from $P[\Delta(t)]$,
and vice versa) must not be affected by global shifts $t\to
t+t_0$ of the readout of the clock. Note that the stochastic
process $\Delta(t)$ need not be stationary, despite the
stationarity of the process $\alpha(t)$.

The one-point probability distribution function for the
variables $\alpha(t)$ should be highly concentrated around the
zero mean, if the clock is to behave nicely, i.e.,
\begin{equation}
\langle\alpha(t)\alpha(t-\tau)\rangle\equiv c(\tau)\leq c(0)\ll
1 \,,
\end{equation}
where $c(\tau)=c(-\tau)$.

The correlation time $\vartheta$ for the stochastic process
$\alpha(t)$ is given by
\begin{equation}
\vartheta=\int_{0}^{\infty}c(\tau)/c(0)\,.
\end{equation}
For convenience, let us introduce a new parameter $\kappa$ with
dimensions of time, defined as $\kappa^2=c(0)\vartheta^2$ and
for which the good-clock conditions imply $\kappa\ll\vartheta$.
As we shall see, $\vartheta$ cannot be arbitrarily large, and,
therefore, the ideal clock limit is given by $\kappa\to0$.

In addition to these properties, a good clock must have enough
precision in order to measure the evolution of the specific
system, which imposes further restrictions on the clock. On the
one hand, the characteristic evolution time $l$ of the system
must be much larger than the correlation time $\vartheta$ of the
clock. On the other hand, the leading term in the asymptotic
expansion of the variance $\langle\Delta(t)^2\rangle$ for large
$t$ is of the form $\kappa^2 t/\vartheta$ which means that,
after a certain period of time, the absolute errors can be too
large. The maximum admissible standard deviation in $\Delta(t)$
must be at most of the same order as $l$. Then the period of
applicability of the clock to the system under study, i.e., the
period of clock time during which the errors of the clock are
smaller than the characteristic evolution time of the system is
approximately equal to $l^2\vartheta/\kappa^2$. For a good
clock, $\kappa\ll\vartheta\ll l$, as we have seen, so that the
period of applicability is much larger than the characteristic
evolution time $l$.

\subsection{Evolution laws}
 We shall now obtain the evolution
equation for the density matrix of an arbitrary quantum system
in terms of the clock time $t$. Let $H$ be the time-independent
Hamiltonian of the system and $S$ its action in the absence of
errors.

For any given realization of the stochastic process $\alpha(t)$
that characterizes a good clock, we can write the density matrix
at the time $t$, $\rho_\alpha(t)$, in terms of the initial
density matrix $\rho(0)$ as
\begin{equation}
\rho_\alpha(t)=\sso_\alpha(t)\cdot \rho(0)\,,
\end{equation}
where the density matrix propagator $\sso_\alpha(t)$ has the
form
\begin{equation}
\sso_\alpha(t)=\int {\cal D}q {\cal D}q'
e^{iS_\alpha[q;t]-iS_\alpha[q';t]}\,.
\end{equation}
Here, $S_\alpha[q;t]=S[q;t]-\int_0^t ds\alpha(s) H[q(s)]$ is the
action of the system for the given realization of the stochastic
process $\alpha(t)$.

The average of the density matrix $\rho_\alpha(t)$ can be
regarded as the density matrix of the system $\rho(t)$ at the
clock time $t$:
\begin{equation}
\rho(t)=\int {\cal D}\alpha {\cal P}[\alpha] \sso_\alpha(t)
\cdot \rho(0)\,.
\end{equation}
In the good-clock approximation, only the two-point correlation
function $c(\tau)$ is relevant, so that we can write the
probability functional as a Gaussian distribution. The
integration over $\alpha(t)$ is then easily performed to obtain
the influence action ${\cal W}$
\begin{equation}
{\cal W}[q,q';t]=-\frac{1}{2}\int_0^t ds\int_0^s ds'
\{H[q(s)]-H[q'(s)]\}c(s-s')\{H[q(s')]-H[q'(s')]\}\,.
\end{equation}
We see that there is no dissipative term there as could be
expected from the fact that the noise source is classical
\cite{fv63,fh65}. Moreover, as the interaction term is
proportional to $H$, there is no response of the system to the
outside noise, which means that the associated impedance is
infinite \cite{callen51,ga91,mandel95}.

Therefore, we see that the effect of using good real clocks for
studying the evolution of a quantum system is the appearance of
an effective interaction term in the action integral which is
bilocal in time. This can be understood as the first term in a
multilocal expansion, which corresponds to the weak-field
expansion of the probability functional around the Gaussian
term. This nonlocality in time admits a simple interpretation:
correlations between relative errors at different instants of
clock time can be understood as correlations between clock-time
flows at those clock instants. The clock-time flow of the system
is governed by the Hamiltonian and, therefore, the correlation
of relative errors induces an effective interaction term,
generically multilocal, that relates the Hamiltonians at
different clock instants.

From the form of the influence action, it is not difficult to
see that, in the Markov approximation and provided that the
system evolves for a time smaller than the period of
applicability of the clock, the density matrix $\rho(t)$
satisfies the master equation
\begin{equation}
\dot\rho(t)=-i\big[H,\rho(t)\big]-(\kappa^2/\vartheta)
\big[H,\big[H,\rho(t)\big]\big]\,,
\end{equation}
where the overdot denotes derivative with respect to the clock
time $t$. Notice that, in the ideal clock limit, $\kappa\to0$,
the unitary von Neumann equation is recovered. We should also
point out that irreversibility appears because the errors of the
clock cannot be eliminated once we have started using it.

From a different point of view, the clock can be effectively
modeled by a thermal bath, with temperature $T_{\rm b}$ to be
determined, coupled to the system. Let $H+H_{\rm int}+H_{\rm b}$
be the total Hamiltonian, where $H$ is the free Hamiltonian of
the system and $H_{\rm b}$ is the Hamiltonian of a bath that
will be represented by a collection of harmonic oscillators
\cite{ga91,mandel95}. The interaction Hamiltonian will be of the
form $H_{\rm int}=\xi H$, where the noise operator $\xi$ is
given by
\begin{equation}
\xi(t)=\frac{i}{\sqrt{2\pi}}\int_0^\infty d\omega
\chi(\omega)
[ a^{\dag}(\omega) e^{i\omega t}-a(\omega) e^{-i\omega t}]\,.
\end{equation}
In this expression, $a$ and $a^{\dag}$ are, respectively, the
annihilation and creation operators associated with the bath,
and $\chi(\omega)$ is a real function, to be determined, that
represents the coupling between the system and the bath for each
frequency $\omega$.

Identifying, in the classical noise limit, the classical
correlation function of the bath with $c(\tau)$, the suitable
coupling between the system and the bath is given by the
spectral density of fluctuations of the clock:
\begin{equation}
T_{\rm b}\chi(\omega)^2=\int_{0}^\infty d\tau
c(\tau)\cos(\omega\tau)\,.
\end{equation}
With this choice, the master equation for evolution according to
real clocks is identical to the master equation for the system
obtained by tracing over the effective bath.

\subsection{Loss of coherence}
The master equation contains a diffusion term and will therefore
lead to loss of coherence. However, this loss depends on the
initial state. In other words, there exists a pointer basis
\cite{zurek81,zurek82,zurek94}, so that any density matrix which
is diagonal in this specific basis will not be affected by the
diffusion term, while any other will approach a diagonal density
matrix. The stochastic perturbation $\alpha(t)H$ is obviously
diagonal in the basis of eigenstates of the Hamiltonian, which
is therefore the pointer basis: the interaction term cannot
induce any transition between different energy levels. The
smallest energy difference provides the inverse of the
characteristic time for the evolution of the system $l$ and,
therefore, the decay constant is $\kappa^2/\vartheta l^2$, equal
to the inverse of the period of applicability of the clock. By
the end of this period, the density matrix will have been
reduced to the diagonal terms and a much diminished remnant of
those off-diagonal terms with slow evolution. In any case, the
von Neumann entropy grows if the density matrix is not initially
diagonal in the energy basis.

The effect of decoherence due to errors of real clocks does not
only turn up in the quantum context. Consider for instance a
classical particle with a definite energy moving under a
time-independent Hamiltonian $H$. Because of the errors of the
clock, we cannot be positive about the location of the particle
in its trajectory on phase space at our clock time $t$.
Therefore we have an increasing spread in the coordinate and
conjugate momentum over the trajectory. For a generic system,
this effect is codified in the classical master equation
\begin{equation}
\dot\varrho=\big\{H,\varrho\big\}+
(\kappa^2/\vartheta)\big\{H,\big\{H,\varrho\big\}\big\}\,,
\end{equation}
where $\varrho(t)$ is the probability distribution on phase
space in clock time. Finally, it should be observed that the
mechanism of decoherence is neither tracing over degrees of
freedom, nor coarse graining, nor dephasing
\cite{giulini96,cooper97}. Even though there is no integration
over time introduced here by fiat, as happens in dephasing in
quantum mechanics, the spread in time due to the errors of the
clock has a similar effect, and produces decoherence.

\subsection{Real clocks and spacetime foam}
As we have seen, there exist strong similarities between the
evolution in spacetime foam and that in quantum mechanics with
real clocks. In both cases, the fluctuations are described
statistically and induce loss of coherence. However, there are
some major differences. In the case of real clocks, the
diffusion term contains only the Hamiltonian of the system
while, in the spacetime foam analysis, a plethora of
interactions appeared. Closely related to this, fluctuations of
the real clock affect in very similar ways to both classical and
quantum evolution; this is not the case in spacetime foam. The
origin of these differences is the nature of the fluctuations
that we are considering and, more specifically, the existence or
not of horizons. Indeed, when studying real clocks, we have
ensured that they satisfied Galilean causality, i.e., that the
real-time parameter always grows as compared with the ideal
time, so that no closed timelike curves are allowed in Galilean
spacetime, whichever clock we are using. This requirement is in
sharp contrast with the situation that we find in spacetime
foam, where we have to consider topological fluctuations that
contain horizons (virtual black holes, time machines, etc.).
Scattering processes in a spacetime with horizons are
necessarily of quantum nature. A classical scattering process in
the presence of these horizons would inevitably lead to loss of
probability because of the particles that would fall inside the
horizons and would never come out to the asymptotic region.

In other words, the underlying dynamics is completely different
in both cases. Spacetime foam provides a non-Hamiltonian
dynamics since the underlying manifold is not globally
hyperbolic. On the other hand, in the case of quantum mechanics
according to clocks subject to small errors, the underlying
evolution is purely Hamiltonian, although the effective one is
an average over all possible Hamiltonian evolutions and becomes
nonunitary.


\section{Conclusions}
\indent

Quantum fluctuations of the gravitational field may well give
rise to the existence of a minimum length in the Planck scale
\cite{ga95}. This can be seen, for instance, by making use of
the fact that measurements and vacuum fluctuations of the
gravitational field are extended both in space and time and can
therefore be treated with the techniques employed for continuous
measurements, in particular the action uncertainty principle
\cite{me92}. The existence of this resolution limit spoils the
metric structure of spacetime at the Planck scales and opens a
doorway to nontrivial topologies \cite{mt73}, which will not
only contribute to the path integral formulation but will also
dominate the Planck scale physics thus endowing spacetime with a
foamlike structure \cite{wh57} with very complicated topology.
Indeed, at the Planck scale, both the partition function and the
density of topologies seem to receive the dominant contribution
from topological configurations with very high Betti numbers
\cite{ha78,1ca97}.

Spacetime foam may leave its imprint in the low-energy physics.
For instance, it can play the role of a universal regulator for
both the ultraviolet \cite{crane86} and infrared \cite{magnon88}
divergences of quantum field theory. It has also been proposed
as the key ingredient in mechanisms for the vanishing of the
cosmological constant \cite{1ca97,coleman88b}. Furthermore, it
seems to induce loss of coherence \cite{ha82} in the low-energy
quantum fields that propagate on it as well as mode-dependent
energy shifts \cite{1ga98}. In order to study some of these
effects in more detail, we have built an effective theory in
which spacetime foam has been substituted by a fixed classical
background plus nonlocal interactions between the low-energy
fields confined to bounded spacetime regions of nearly Planck
size \cite{1ga98}. In the weak-coupling approximation, these
nonlocal interactions become bilocal. The low-energy evolution
is nonunitary because of the absence of a nonvanishing timelike
Hamiltonian vector field. The nonunitarity of the bilocal
interaction can be encoded in a quantum noise source locally
coupled to the low-energy fields. From the form of the influence
functional that accounts for the interaction with spacetime
foam, we have derived a master equation for the evolution of the
low-energy fields which contains a diffusion term, a damping
term, and energy shifts that can be interpreted as gravitational
Lamb and Stark effects. We have also discussed the size of these
effects as well as the possibility of observing them in the near
future.

We have seen that the evolution of quantum systems according to
good real clocks \cite{egusquiza98} is quite similar to that in
spacetime foam. Indeed, we have argued that good classical
clocks, which are naturally subject to fluctuations, can be
described in statistical terms and we have obtained the master
equation that governs the evolution of quantum systems according
to these clocks. This master equation is diffusive and produces
loss of coherence. Moreover, real clocks can be described in
terms of effective interactions that are nonlocal in time.
Alternatively, they can be modeled by an effective thermal bath
coupled to the system. In view of this analysis, we have seen
that, although there exist strong similarities between
propagation in spacetime foam and according to real clocks,
there are also important differences that come from the fact
that the underlying evolution laws for spacetime foam are
nonunitary because of the presence of horizons while, in the
case of real clocks, the underlying evolution is unitary and the
loss of coherence is due to an average over such Hamiltonian
evolutions.

\section*{Acknowledgments}
\indent

I am grateful to C. Barcel\'{o} and P.F. Gonz\'{a}lez-D\'{\i}az for helpful
discussions and reading the manusscript. I was supported by
funds provided by DGICYT and MEC (Spain) under Projects
PB93--0139, PB94--0107, and PB97--1218.


\end{document}